\documentclass[reprint, pra, superscriptaddress]{revtex4-1}
\usepackage{color,amssymb}
\usepackage{graphicx}
\usepackage{float}

\newcommand{\jkg}[1]{\textcolor{black}{#1}}
\newcommand{\tledit}[1]{\textcolor{black}{#1}}
\newcommand{\jck}[1]{\textcolor{black}{#1}}
\newcommand{\new}[1]{\textcolor{black}{#1}}

\begin{document}


\title{Atomic-layer doping of SiGe heterostructures for atomic-precision donor devices } 



\author{E. Bussmann}
\email[]{ebussma@sandia.gov}
\affiliation{Center for Integrated Nanotechnologies, Sandia National Laboratories, Albuquerque NM 87185, USA}

\author{\jkg{John King} Gamble}
\email{\jkg{john.gamble}@sandia.gov}
\affiliation{Center for Computing Research, Sandia National Laboratories, Albuquerque, NM 87185, USA}

\author{J. \jck{C.} Koepke}
\affiliation{Sandia National Laboratories, Albuquerque NM 87185, USA}

\author{D. Laroche}
\affiliation{Sandia National Laboratories, Albuquerque NM 87185, USA}

\author{S. H. Huang}
\affiliation{Department of Electrical Engineering and Graduate Institute of Electronic Engineering, National Taiwan University, Taipei 10617, Taiwan, R.O.C.}
\affiliation{National Nano Device Laboratories, Hsinchu 30077, Taiwan, R.O.C.}

\author{Y. Chuang}
\affiliation{Department of Electrical Engineering and Graduate Institute of Electronic Engineering, National Taiwan University, Taipei 10617, Taiwan, R.O.C.}
\affiliation{National Nano Device Laboratories, Hsinchu 30077, Taiwan, R.O.C.}

\author{J.-Y. Li}

\affiliation{Department of Electrical Engineering and Graduate Institute of Electronic Engineering, National Taiwan University, Taipei 10617, Taiwan, R.O.C.}
\affiliation{National Nano Device Laboratories, Hsinchu 30077, Taiwan, R.O.C.}

\author{C. W. Liu}

\affiliation{Department of Electrical Engineering and Graduate Institute of Electronic Engineering, National Taiwan University, Taipei 10617, Taiwan, R.O.C.}
\affiliation{National Nano Device Laboratories, Hsinchu 30077, Taiwan, R.O.C.}

\author{B. S. Swartzentruber}
\affiliation{Center for Integrated Nanotechnologies, Sandia National Laboratories, Albuquerque NM 87185, USA}

\author{M. P. Lilly}
\affiliation{Center for Integrated Nanotechnologies, Sandia National Laboratories, Albuquerque NM 87185, USA}

\author{M. S. Carroll}
\affiliation{Sandia National Laboratories, Albuquerque NM 87185, USA}

\author{T.-M. Lu}
\affiliation{Sandia National Laboratories, Albuquerque NM 87185, USA}


\date{\today}

\begin{abstract}
As a first step to porting scanning tunneling \jkg{microscopy} methods of atomic-precision fabrication to a strained-Si/SiGe platform, we demonstrate post-growth P atomic-layer doping of SiGe heterostructures. To preserve the substrate structure and elastic state, we use a T $\leq 800^\circ$C process to prepare clean Si$_{0.86}$Ge$_{0.14}$ surfaces suitable for atomic-precision fabrication. P-saturated atomic-layer doping is incorporated and capped with epitaxial Si under a thermal budget compatible with atomic-precision fabrication. Hall measurements at T$=0.3$ K show that the doped heterostructure has R$_{\square}=570\pm30$ $\Omega$, yielding an electron density $n_{e}=2.1\pm0.1\times10^{14}$cm$^{-2}$ and mobility $\mu_e=52\pm3$ cm$^{2}$ V$^{-1}$ s$^{-1}$, similar to saturated atomic-layer doping in pure Si and Ge. The magnitude of $\mu_e$ and the complete absence of Shubnikov-de Haas oscillations in  magnetotransport measurements indicate that electrons are overwhelmingly localized in the donor layer, and not within a nearby buried Si well. This conclusion is \jkg{supported} by self-consistent \jkg{Schr\"odinger-Poisson} calculations that predict electron occupation primarily in the donor layer.
 \end{abstract}


\maketitle 


\section{Introduction}

Atomic-precision fabrication via scanning tunneling microscopy (STM) hydrogen depassivation lithography has blossomed as a unique route to form doped-Si and Ge nanoelectronics with single-atom selectivity and feature sizes~\cite{r04, s04, f7,f12,gs9,gs11}.  
Atomic-precision is enabling for Si quantum computing research~\cite{z12,K98,Vr00,Fang05, H6, H15, k01, JK15} with potential utility in other fields such as limit-testing in scaling and quantum confined geometries in conventional microelectronics~\cite{k11,P7,W14,Sc17,gr17}. 

A step to facilitate broader electronic device applications is to integrate atomic-precision doping into vertically-gated configurations incorporating metal-insulator-semiconductor (MIS) capacitors~\cite{S7, MF7,L10}. \jkg{For example, the archetypal Kane proposal utilizes} single donors placed in MIS stacks~\cite{K98,Vr00,Fang05, H6}. To protect atomic-precision donors from disordering via diffusion, the thermal budget to integrate a gate stack is constrained to temperatures too low for ideal SiO$_2$ growth by furnace oxidation~\cite{g9,s4,k15}. This has inspired attempts to integrate gate stacks using insulators grown by lower-\tledit{temperature} methods. Unfortunately, these yield more defective films and disorder with measurable effects on donor device performance~\cite{MF7,L10}.

\begin{figure}
 \includegraphics[width=200 pt]{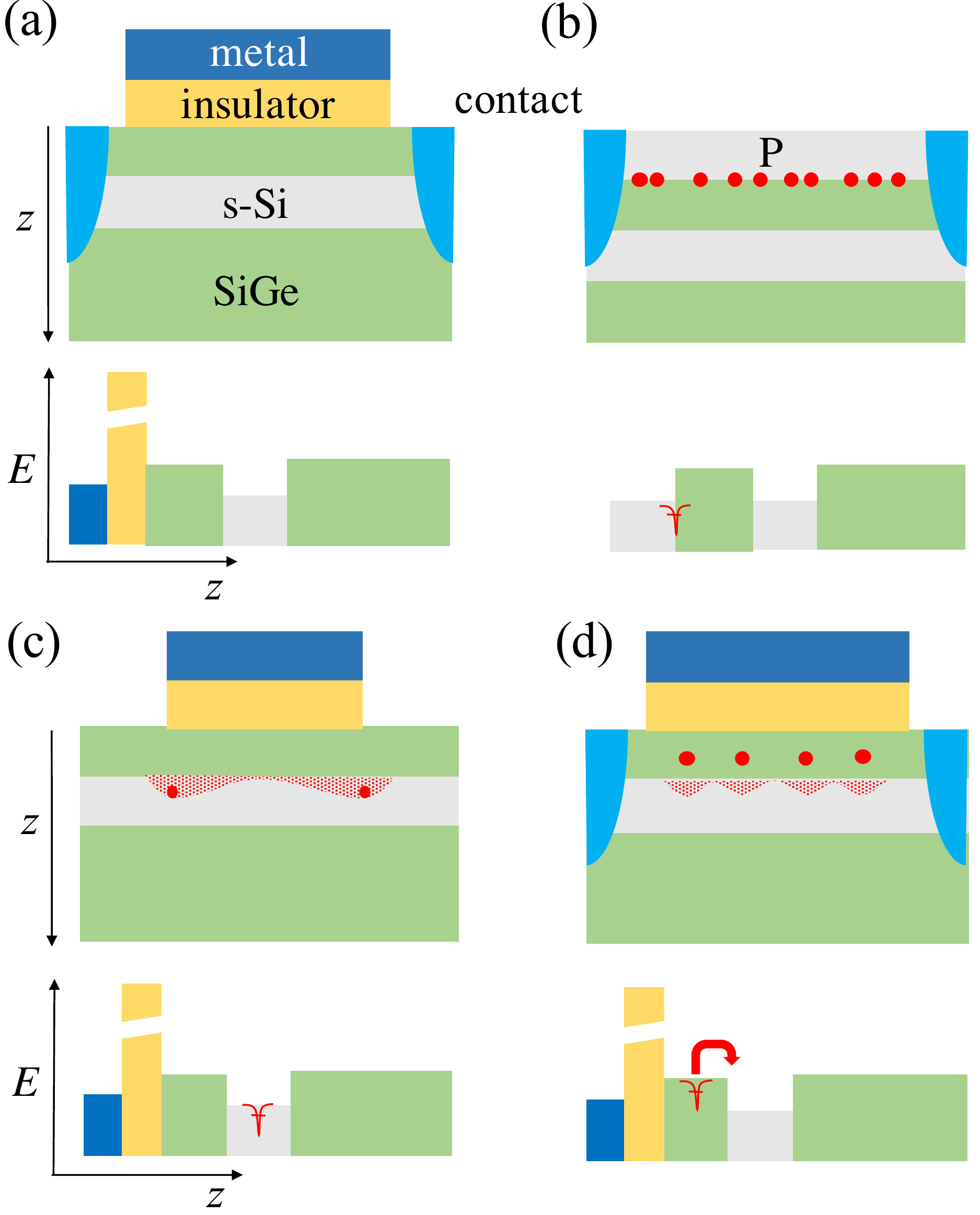}%
\caption{Schematic cross sections of heterostructure test devices and proposed device concepts. The lower panels show schematics of relative conduction band position in each layer. (a) An undoped field-effect transistor for characterizing transport of gate-accumulated two-dimensional electrons in the s-Si well prior to our atomic-layer doping process. (b) A schematic of the atomic-layer-doped transport device produced in this work. The doping is a two-dimensional (2D) random substitutional Si:P alloy. Electron transfer from the donor layer to the s-Si well is blocked by the band alignments. (c) Gate-stack isolation for donor spin qubits via band offsets in an all-epitaxial Si environment away from Si-insulator interfaces. Spin initialization and readout would be enabled by gate-defined accumulation-mode single-electron transistors (not shown). (d) Concept for a high electron mobility transistor where STM-defined atomic-precision modulation doping creates a periodic potential in the s-Si transport channel.}%
\label{fig:STACK} 
\end{figure}

Strained-layer band-offset engineering via thin-film heteroepitaxy is a mature technique to form an interfacial barrier at a relatively low growth thermal budget ~\cite{ML5,Vr00}. Layered strained-Si (s-Si) and SiGe heterostructures have the benefit of pristine all-epitaxial interfaces \new{with disorder that compares well with the highest-quality Si/SiO$_2$ interfaces ~\cite{ML5,F3}. As a measure of disorder, exceptional Si/SiO$_2$ field-effect devices undergo a $T<1$~K  metal insulator transition at critical electron densities $n_c=0.6$-$0.8\times10^{11}$ cm$^{-2}$ with peak mobilities $\mu_e\sim2$-$5\times10^{4}$ cm$^2$V$^{-1}$s$^{-1}$ ~\cite{P93, K17}. By comparison, good SiGe field-effect devices turn on at $n_c=0.3$-$0.6\times10^{11}$ cm$^{-2}$ achieving peak $\mu_e\sim10^{5}$ -$10^6$ cm$^2$V$^{-1}$s$^{-1}$~\cite{L05,l9,L15, XMi15, Z15}. In our SiGe substrate, we find $n_c=0.6\times10^{11}$ cm$^{-2}$ and peak $\mu_e = 7\times10^{5}$ cm$^2$V$^{-1}$s$^{-1}$ at T$=0.3$~K prior to adding atomic-layer doping~\cite{L15}.  
Fig.~\ref{fig:STACK} (a) shows a schematic of the field-effect device used to characterize transport in the s-Si well (see Ref.~\tledit{\onlinecite{L15}}).  Fig.~\ref{fig:STACK} (b) shows our device to test transport after atomic-layer doping.}

Integrating atomic-precision doping with SiGe heterostructures leverages advantages of all-epitaxial order inherent in both platforms. A few works have proposed donor spin qubits in s-Si/SiGe ~\cite{Vr00, Fang05}. Fig.~\ref{fig:STACK} (c) shows a simplified sketch for a 2-qubit interaction. Qubit control would be via MIS gating to move electrons \jkg{on and off} the nucleus, or into SiGe layers, to tune spin precession. Spins would couple by drawing electrons to heterointerfaces, {\it e.g.}, to form 2-electron quantum dots. Spin initialization and readout would be by established methods using single-electron transistors formed by MIS accumulation in the well~\cite{Vr00,Fang05}. Another potential application is 2D atomic-precision modulation doping to investigate effects of carefully tuned ordered potentials on 2D electron transport at heterostructure interfaces, which may permit tailor-made electronic band structures enforced by donor periodicity~\cite{k11}. Fig~\ref{fig:STACK} (d) shows a schematic of a field-effect device with ordered modulation doping in the channel. Ordered doping potentials promote channel mobility, with anticipated applications in high-electron mobility transistors~\cite{ML5}.

Here,  we demonstrate atomic-layer doping of a s-Si/SiGe heterostructure via a thermal process entirely compatible with STM atomic-precision fabrication and  our epitaxial SiGe heterostructure, which can withstand T $\lesssim$850$^{\circ}$C annealing~\cite{Kl96}.  We incorporate P atomic-layer doping into the heterostructure by a process thermally compatible with atomic-precision fabrication. Results of each process step are characterized and contrasted with outcomes on relaxed bulk Si. Post-growth secondary ion mass spectroscopy (SIMS) yields an integrated P density, $N_D\sim 1.2\times 10^{14}$ cm$^{-2}$ with a \tledit{full width at half maximum (FWHM)}$=$3 nm, similar to results demonstrated for pure Si and Ge~\cite{O2,gs9}. Hall effect measurements show that the doped heterostructure has an electron density, $n_e=2.1\pm 0.1\times10^{14} $cm$^{-2}$ and mobility $\mu_e =52\pm3$ cm$^2$V$^{-1}$s$^{-1}$. These numbers are consistent with electron transport predominantly in the atomic-layer doping. Magnetotransport measurements to $\pm$8 T reveal only a small B$=$0 T weak-localization bump, characteristic of transport via the atomic-layer doping,  and no sign of Shubnikov-de Haas oscillations characteristic of transport via the high-mobility strained Si well~\cite{L15}. Self-consistent \jkg{Schr\"odinger-Poisson} calculations confirm the intuitive picture that electrons primarily populate the donor layer.

\section{Device fabrication and physical characterization}

Atomic-precision fabrication has been demonstrated on (100)-oriented Si~\cite{r04, s04, f7,f12}, Ge~\cite{gs9,gs11}, strained-Si on insulator (s-SOI)~\cite{l13, l14}, and Ge on insulator ~\cite{k13}. Needless to say, the recipes differ between each, owing to unique thermal and chemical considerations, {\it cf} materials-specific growth recipes for molecular beam epitaxy (MBE) of Si vs Ge~\cite{i86,O98}. In principle, s-SOI also provides a platform for strained-layer band engineering, but we consistently find that the s-SOI surface is rough on a nanometer-scale owing to bulk-elasticity-driven step-bunching~\cite{l13,l14,T95}. Therefore, we have chosen to investigate donor placement on the relaxed SiGe surface, which as we will show, does not suffer from the inherent step bunching problem.

We perform atomic-layer doping in an ultrahigh vacuum (5$\times10^{-10}$ Torr) scanning tunneling microscope (UHV STM) equipped with a phosphine (PH$_3$) gas source and a Si \tledit{MBE} source (joule heated solid Si, homebuilt). The substates for our experiments are s-Si/Si$_{0.86}$ Ge$_{0.14}$ heterostructures grown by UHV chemical vapor deposition using SiH$_4$ and GeH$_4$ as precursors at T$=$550 $^{\circ}$C. The substrates for our experiments have a 20-nm-thick s-Si well and a 50-nm-thick relaxed Si$_{0.86}$Ge$_{0.14}$ cap. Further details about the substrate growth and characterization can be found in \jkg{Ref}.~\tledit{\onlinecite{L15}}.

\begin{figure*}
 \includegraphics[width=380 pt]{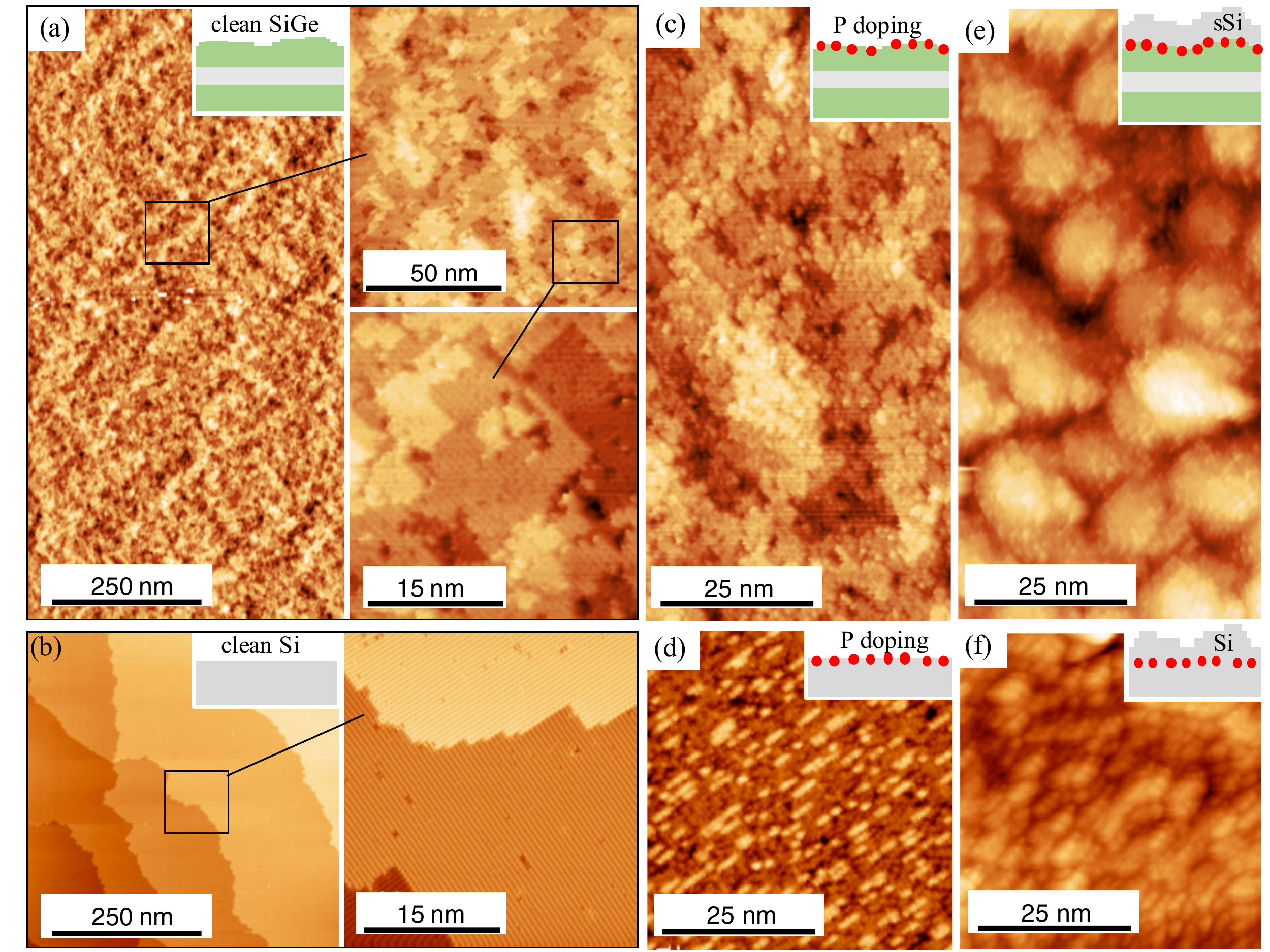}
\caption{ STM images of the outcomes of the atomic-layer doping process on SiGe along with images of the process on the bulk Si(100) surface for comparison. (a) STM of Si$_{0.86}$Ge$_{0.14}$ surface following surface preparation by 650 $^{\circ}$C/15 s annealing in UHV. Smaller panels show details of the surface dimer-row reconstruction and monolayer-sized roughness. (b) A bulk Si surface prepared by a 1200 $^{\circ}$C/10 s anneal. (c) SiGe surface following P incorporation and (d) a bulk Si surface after P incorporation covered by monolayer-high islands of Si ejected by P incorporation into the surface. (e) SiGe:P after capping by 20-nm-thick epitaxial s-Si, and (f) the surface of a 20-nm-thick epitaxial Si cap on a Si:P structure. All images were acquired in filled-states at $-2$ to $-4$ V bias and $I=100$ pA.}
\label{fig:STM} 
\end{figure*}

Prior to atomic-layer doping experiments, we perform an {\it ex-situ} wet chemical cleaning procedure, and annealing in the UHV STM.  The samples are cleaned by three cycles of chemical oxidation ($3:1$ H$_2$SO$_4:$H$_2$O$_2$, 90 $^{\circ}$C, 5 mins) and reduction ($6:1$ NH$_4$F:HF, 10 s), followed by a rinse in deionized H$_2$O. Samples are exceptionally hydrophobic upon extraction (presumably hydrogen terminated). Two different chemical and annealing recipes yield suitably clean flat surfaces for atomic layer doping: (1) the H-terminated samples can be loaded into the UHV system and annealed at \tledit{a temperature} sufficient to desorb the hydrogen (650 $^{\circ}$C, 15 s), or (2) samples can be oxidized (5:1:1 H$_2$O:H$_2$O$_2$:HCl, 60 $^{\circ}$C, 5 mins), then annealed hotter (T$=$800 $^{\circ}$C, 20 mins) to remove the chemical oxide. Both approaches yield suitable surfaces and there is not yet any evident advantage to either approach.

Fig.~\ref{fig:STM} (a) shows STM images of a SiGe(100) surface prepared by our {\it ex-situ} chemical process followed by UHV annealing at T$=$650 $^{\circ}$C for 15 s. For contrast, Fig.~\ref{fig:STM} (b) shows images of pure Si(100) surfaces prepared by annealing at 1200 $^{\circ}$C for 10 s, yielding a characteristic step-terrace structure with the 2$\times$1/1$\times$2 reconstruction.

Our lower-\tledit{temperature} process produces surfaces that show no definitive indication of any step-terrace order in larger areas,  Fig.~\ref{fig:STM} (a), but higher magnification images (insets) reveal that the surface is simply rough at the atomic-layer scale \jck{with terraces only several nanometers wide. The terraces have mixed \new{p($2\times2$) and c($4\times2$)} reconstructions characteristic for SiGe and pure Ge~\cite{gs9}.} The atomic-layer roughness is most likely caused by the chemical preparations, and the subsequent anneal is inadequate to activate surface smoothing.  Roughness at the atomic-layer scale does not cause problems for atomically-precise donor fabrication. Recently, McKibbin {\it et al.} have demonstrated entirely functional atomically-precise donor devices on similarly rough epitaxial Si surfaces~\cite{m9,m13}. 

Atomic-layer doping is introduced by \new{exposing the sample} to PH$_3$ at  2$\times10^{-8}$ Torr for 5 mins (6 Langmuir dose) at room temperature. A T$=$400 $^{\circ}$C/30 s anneal is used to incorporate P into \new{lattice sites in the surface}~\cite{r04,s04,l9,l13,k13}. 
Fig.~\ref{fig:STM} (c) shows the surface after the incorporation anneal, which again is not hot enough to activate surface smoothing, leaving the characteristic atomic-layer roughness on the surface. For comparison, Fig.~\ref{fig:STM} (d)  shows an image of a relaxed bulk Si surface after a similar doping procedure. The surface is covered by $\sim 0.25$ monolayers of atomic-height islands formed by Si ejected onto the surface by P incorporation~\cite{r04}. We do not observe similar islands on the SiGe surface most likely because ejected Si or Ge is able to migrate to dense preexisting atomic steps, preventing island nucleation.

After P incorporation, samples are capped with 10 to 20 nm thick \new{epitaxial s-Si} grown at $\sim0.1$ \AA/s at T$=$250 $^{\circ}$C. Previous reports for Si and Ge structures indicate that this temperature is sufficiently low to inhibit P surface segregation, while yielding a crystalline capping layer~\cite{O2}. Fig.~\ref{fig:STM} (e)  shows an STM image of the surface of the capping layer, which is rough owing to the relatively low growth temperature. Under similar growth conditions, similarly rough surfaces are reported for both Si and Ge epitaxy~\cite{r04,gs9}. Fig.~\ref{fig:STM} (f)  shows an example of an STM image of the typical surface topography of a capping layer on pure bulk Si. 

\begin{figure}
 \includegraphics[width=210pt]{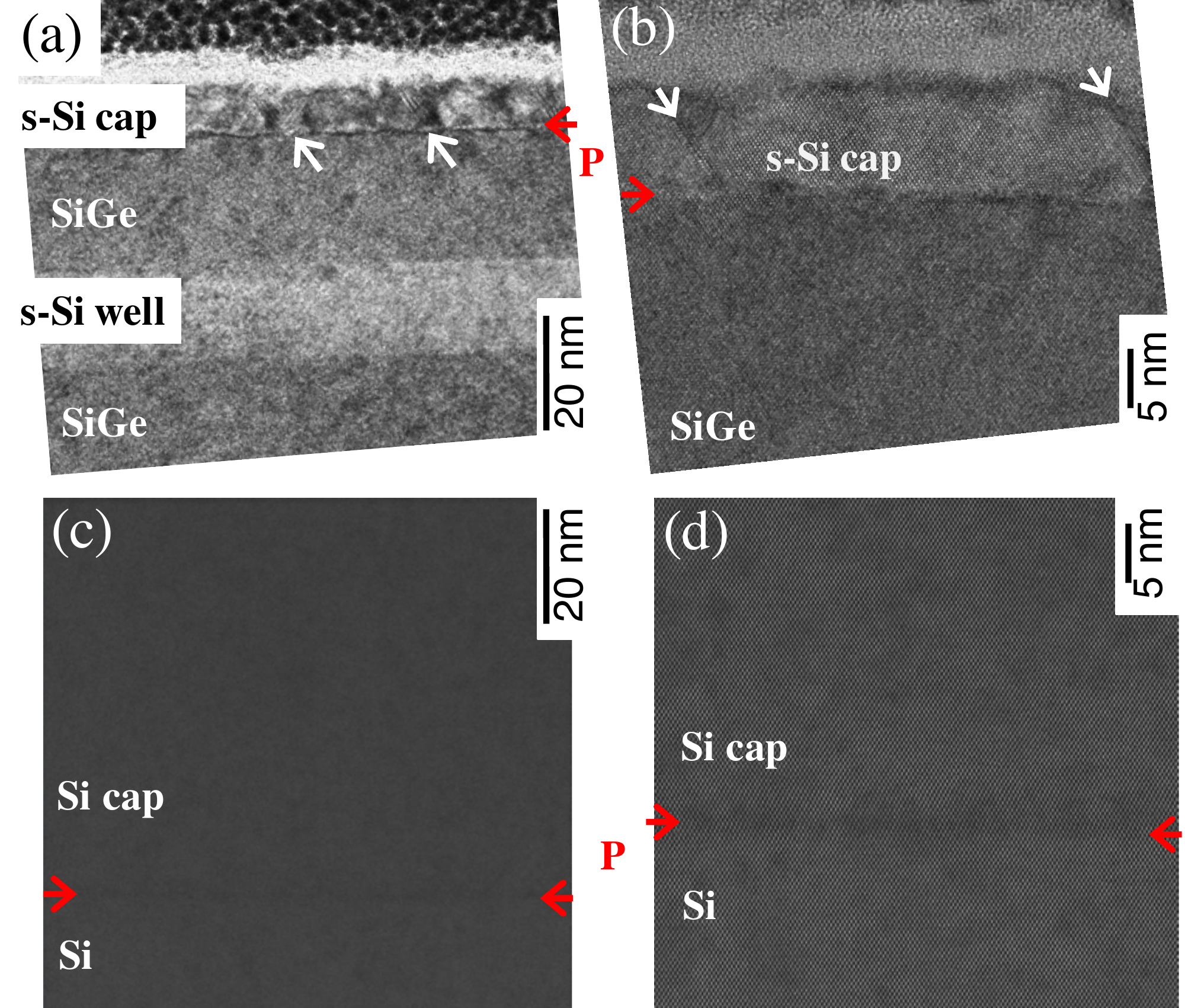}%
\caption{ Cross-sectional TEM of atomic-layer doped structures: (a) s-Si/P/Si$_{0.86}$Ge$_{0.14}$, (b) detail of the s-Si/P/Si$_{0.86}$Ge$_{0.14}$ interfaces and crystal, (c) Si:P doped structure grown under similar conditions, and (d) interfacial and crystal detail of the Si:P structure. Red arrows point to the P layer and white arrows point to stacking faults in the epitaxial s-Si cap. } 
\label{fig:TEM}
\end{figure}

Although there is no clear indication of crystal order, e.g. the 2$\times1$ reconstruction, in the STM images of the epitaxial s-Si cap, cross-sectional transmission electron microscopy (TEM), Fig.~\ref{fig:TEM}, shows that the films are epitaxial. \new{Figs.}~\ref{fig:TEM} (a) and (b) show that the SiGe layer structure is preserved  with no evidence of threading dislocations in the field of view. Stacking faults are visible in the s-Si cap on the SiGe material. Figs.~\ref{fig:TEM} (c) and (d) show cross-sectional TEM of the relaxed bulk Si:P structure for comparison. There are no stacking faults in the relaxed Si:P structure.

\begin{figure}
 \includegraphics[width=240 pt]{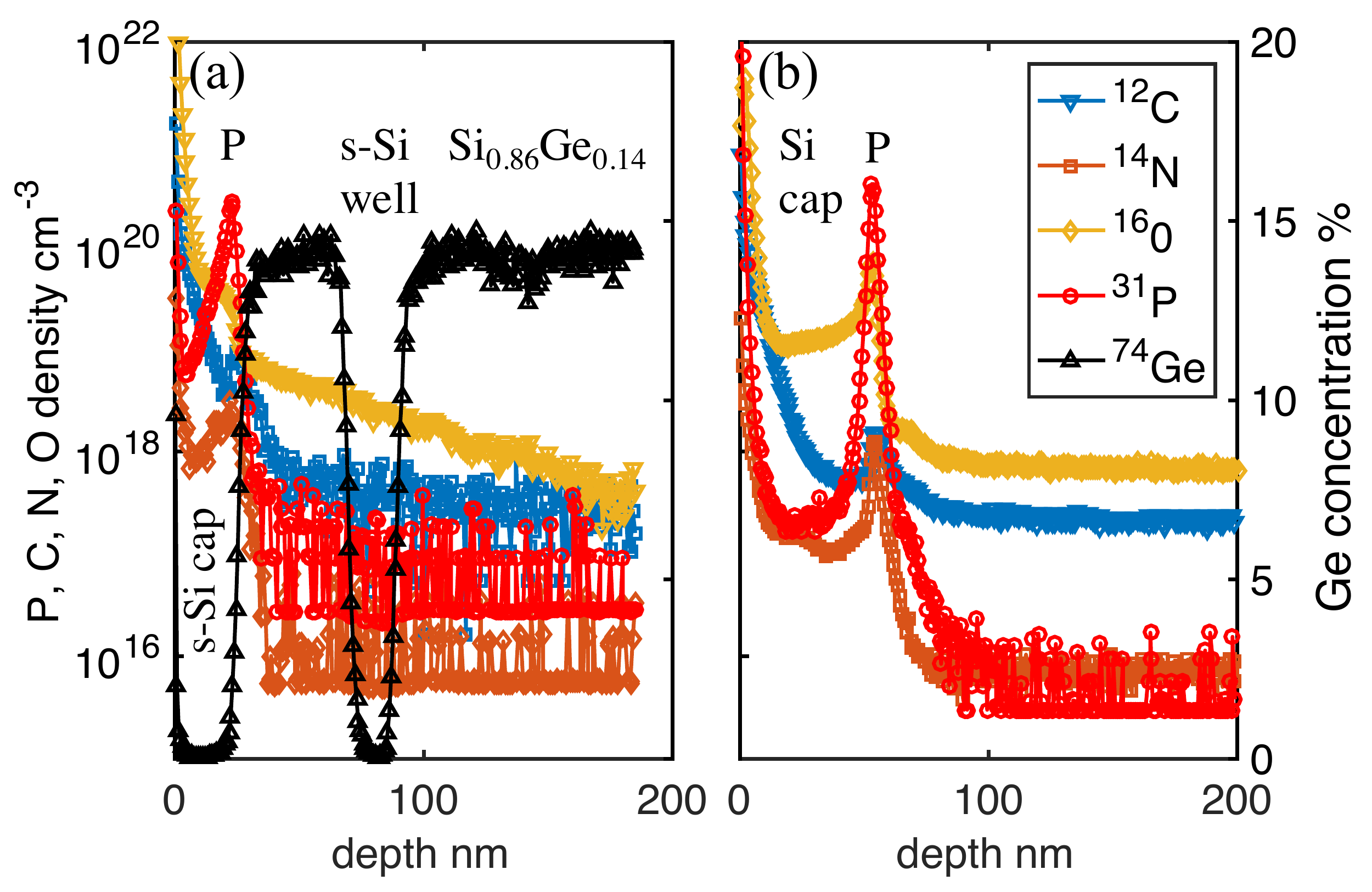}%
\caption{(a) Post-growth SIMS of the complete s-Si/P/Si$_{0.86}$Ge$_{0.14}$.(b) SIMS of Si:P material grown under similar conditions. }%
\label{fig:SIMS}
\end{figure}

Secondary ion mass spectroscopy (SIMS), Fig.~\ref{fig:SIMS} (a), shows that our recipe yields a sharply peaked, FWHM $=$ 3 nm, P sheet density $\sim1.2\times 10^{14}$ cm$^{-2}$.  This is very similar to the FWHM$=$3 nm peaked P sheet density 1.3$\times 10^{14}$ cm$^{-2}$, measured in our Si:P structure, Fig.~\ref{fig:SIMS} (b), and consistent with results reported previously~\cite{O2}. The apparent 3-nm-spread of the dopant peak is due to the limited SIMS depth resolution ($\sim$4 nm/decade of P density). The SiGe epitaxial cap has substantial backgrounds of N,C, and O contamination, which are somewhat higher than in the bulk Si:P structure, indicating that the lower-\tledit{temperature} process leaves more residual contamination, which may be the cause of the stacking faults in the s-Si capping layer~\cite{Fi63}.  

\section{Electronic device characterization}
	
\begin{figure}
 \includegraphics[width=220pt]{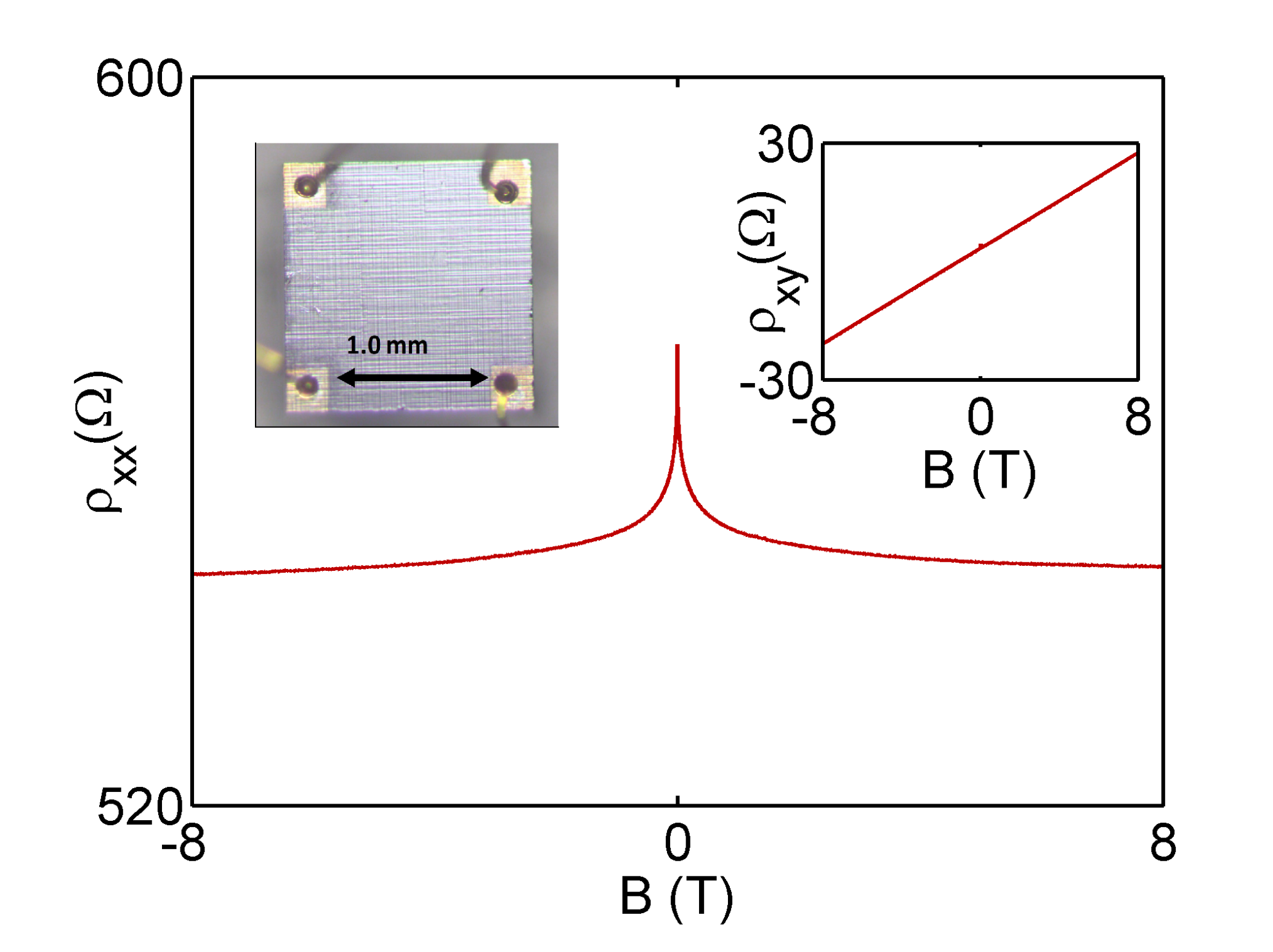}%
\caption{ \tledit{Longitudinial magnetoresistance and Hall (inset) resistance at T$=$0.3 K of the atomic-layer doped heterostructure} along with a photo of the test device. The aluminum contact pads appear bright gold hue.}%
\label{fig:TRANS}
\end{figure}

Electronic figures-of-merit for the atomic-layer doping are the electron density and mobility. These are measured using a Hall effect device fabricated on a $\sim1\times$1 mm$^2$ die with Al contact pads. Fig.~\ref{fig:TRANS} inset shows a photograph of the completed device.  

Transport measurements are done in a $^3$He system with a base temperature of 0.3 K.  We use standard lock-in measurement techniques with an excitation current of 100 nA. Hall effect measurements, Fig.~\ref{fig:TRANS} inset, yield R$_{\square}=570\pm30$ $\Omega$, and an electron density n$_{e}=2.1\pm 0.1\times10^{14}$ cm$^{-2}$ and mobility $\mu_e=52\pm3$ cm$^{2}$V$^{-1}$s$^{-1}$. These values are similar to those reported for atomic-layer doping in  Si (1.7-2.4$\times 10^{14}$ cm$^{-2}$ and $\mu_e =$30-120 cm$^2$V$^{-1}$s$^{-1}$), and Ge (6$\times$10$^{13}$ cm$^{-2}$, 30 cm $^2$V$^{-1}$s$^{-1}$)~\cite{O2,g9,gs9}. These values of mobility also compare well with the calculations of Hwang and Das Sarma describing transport in the donor layers entirely in a semiclassical Drude model~\cite{H13}.

Magnetotransport measurements, Fig.~\ref{fig:TRANS}, provide insight into the character of transport. The data is essentially featureless except for a 4 $\%$ bump at B$=$0 T, indicative of weak localization which is typical of transport via atomic-layer doping in Si and Ge~\cite{O2,g9}. The values of $n_e$, $\mu_e$, and the absence of any Shubnikov-de Haas oscillations indicate that electrons move via the atomic-layer doping, and do not significantly populate the nearby higher-mobility buried Si well (20-50 nm distant).

\section{Self-consistent simulations}

\new{ To determine whether electron transport via the 2D phosphorus sheet is physically consistent with the device geometry,} we performed self-consistent simulations of the \tledit{one-dimensional} semiconductor heterostructure (along the growth direction). 
We use the standard predictor-corrector approach to achieve rapidly-convergent calculations \cite{trellakis:1997}, assuming zero-field boundary \jkg{condition} at the bottom of the stack, \jkg{corresponding to charge neutrality deep in the substrate.} 
\jkg{Since the conduction band is not pinned at the top of the stack (either by a top gate or a high density of surface traps),
the appropriate boundary condition is ambiguous.}
\jkg{Hence, in our simulations we checked a wide range of Dirichlet boundary conditions at the top to ensure that our conclusions do not rely on particular boundary conditions.}
Throughout, we work within the Hartree approximation at 4 K, and neglect exchange and correlation effects.

\jkg{The} non-trivial part of the calculation is the proper inclusion of the P-doped layer, which is sufficiently highly doped that typical models of donor neutralization \cite{gao:2013} no longer directly apply.
Instead, we explicitly include the $\Gamma_1$, $\Gamma_2$, and four degenerate $\Delta$ bands in the 2D density of states model introduced in Ref.~\tledit{\onlinecite{drumm:2012}}, with the locations of the 
P-layer bands specified by Ref.~\tledit{\onlinecite{carter:2009}}.
As pointed out in Ref.~\tledit{\onlinecite{drumm:2012}}, the simple density of states model alone introduces an inconsistency, in that the Fermi level required to impose charge neutrality is predicted to be (unphysically) above
the conduction band edge for 1/4 monolayer P-doping.
To achieve the Fermi level location of $130$ meV below the conduction band edge, as reported from DFT \cite{drumm:2012}, we adopt a uniform multiplicative correction to the density of states of the three P-layer bands.
\jkg{We find they} need to be suppressed by $\sim0.65$.
The P-layer is smeared out over a 2~nm \jkg{vertical} window, and its bands are included alongside conduction band accumulation in our self-consistent simulations.

\begin{figure}[tb]
\includegraphics[width= 1.0 \linewidth]{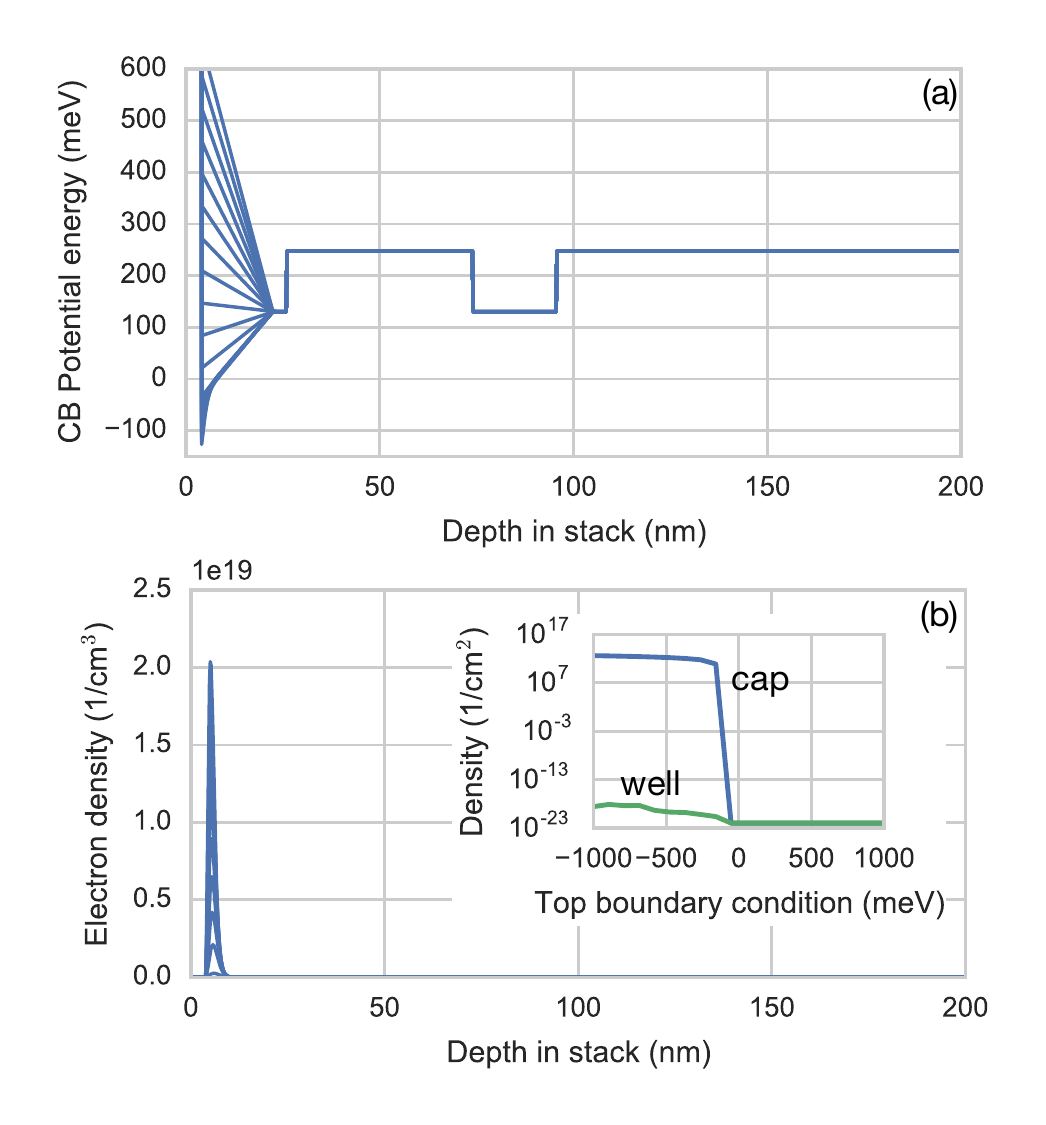}
\caption{\label{fig:devSim} Self-consistent simulations of the atomically-doped structure.
(a), Conduction band edge results from self-consistent \jkg{Schr\"odinger-Poisson} simulations.
\jkg{Since the boundary condition at the top of the stack is uncertain, we computed results for many values, ranging between $\pm 1$~eV.}
\jkg{We found that the} pinning due to the P-layer was sufficiently strong that surface effects (either via gating or oxide charge) would have a negligible effect on the quantum well below.
The conduction band at the P-layer is pinned 130~meV above the conduction band edge, which is the charge-neutrality condition established in Ref.~\tledit{\onlinecite{carter:2009}}.
(b), Excess electron density (\jkg{electrons} neutralizing the P-layer are not shown) accumulation in the heterostructure stack for the various top boundary conditions used.
The inset shows the integrated 2D sheet density of excess electrons in the cap layer and the quantum well, respectively\jkg{, as a function of top boundary condition}.
\jkg{This shows that, although we can accumulate electrons in the cap layer we never expect accumulation within the quantum well.}
 Zero density was set to $1\times10^{-23}$~cm$^{-2}$ for plotting on a semilog scale.
}
\end{figure}

In Fig.~\ref{fig:devSim}, we show results of our calculations. 
As \jkg{can be anticipated from} the experiment, the conduction band is \jkg{effectively} pinned by the P-layer, \jkg{which is incredibly metallic}.
Since the charge-neutrality condition imposes that the conduction band is 130~meV above the Fermi level at the P-layer, essentially no accumulation occurs in the quantum well, which must dip below the Fermi level to form a channel. 
 \jkg{As we discussed above, the top interface is not directly controlled, so} we confirm that a wide variety of boundary conditions do do not change the core results. 
 While \jkg{certain plausible boundary conditions can} achieve modest accumulation in the silicon cap, we cannot form a channel in the quantum well. 

\new{These results indicate that, given our device geometry, we would expect electron occupation in the 2D phosphorus sheet, but not in the quantum well. Hence, our device simulations provide a consistent picture with our electrical transport measurements.}

\section{Summary}	

In summary, we have demonstrated a recipe for low-\tledit{temperature} surface preparation and atomic-layer doping of SiGe heterostructures that is compatible with atomic-precision donor nanofabrication, while preserving the strained-layer structure and elastic state.
SIMS shows the atomic-layer doping has  N$_D\sim 1.2\times 10^{14}$ cm$^{-2}$ with a FWHM$=$3 nm, while Hall effect measurements at T$=$0.3 K show that the structure has a sheet resistance R$_{\square}=570\pm30$ $\Omega$, electron density n$_{e}=2.1\pm 0.1\times10^{14} $cm$^{-2}$ and mobility $\mu_e=52\pm3$ cm$^{2}$V$^{-1}$s$^{-1}$. These results indicate that electrons stay localized to the donors owing to the conduction band offset of the s-Si cap wth respect to the SiGe. By contrast, to achieve spontaneous electron transfer from the donors to the well would require capping the donor layer with SiGe, as in Fig.~\ref{fig:STACK} (d).  Also while we have demonstrated atomic-layer doping on the relaxed Si$_{0.86}$Ge$_{0.14}$ surface, the procedure will work equally well on s-Si surfaces formed by epitaxy on the relaxed SiGe structure~\cite{l13,l14}. This is an enabling step toward engineering structures that simultaneously utilize atomic-precision donor nanofabrication and SiGe strained-layer engineering. Our proof-of-concept is on a complete heterostructure with a preprocess peak mobility $\mu_e=7\times10^5$ cm$^{2}$V$^{-1}$s$^{-1}$, although the procedure will work equally on a commercially available relaxed SiGe virtual substrate. The next step is to demonstrate a complete atomic-precision nanofabrication process using STM hydrogen depassivation lithography to template donor incorporation laterally in-plane to form 3-D atomically abrupt doped nanostructures.~\cite{r04, s04, f7}

\begin{acknowledgments}

We would like to thank Matthew T. Janish and William M. Mook for cross-sectional TEM. Thanks to Stephen M. Carr and Daniel R. Ward for useful discussions and critical reading of this manuscript. This work has been supported by the Division of Materials Sciences and Engineering, Office of Basic Energy Sciences, U.S. Department of Energy (DOE). This work was performed, in part, at the Center for Integrated Nanotechnologies, an Office of Science User Facility operated for the U.S. Department of Energy (DOE) Office of Science. Sandia National Laboratories is a multimission laboratory managed and operated by National Technology and Engineering Solutions of Sandia, LLC., a wholly owned subsidiary of Honeywell International, Inc., for the U.S. Department of Energy's National Nuclear Security Administration under contract DE-NA-0003525. The work at NTU has been supported by the Ministry of Science and Technology (103-2112-M-002-002-MY3 and 105-2622-8-002-001) and National Applied Research Laboratories through I-Dream project (05105A2120).

\end{acknowledgments}

\bibliography{REFS}

\begin{thebibliography}{50}%
\makeatletter
\providecommand \@ifxundefined [1]{%
 \@ifx{#1\undefined}
}%
\providecommand \@ifnum [1]{%
 \ifnum #1\expandafter \@firstoftwo
 \else \expandafter \@secondoftwo
 \fi
}%
\providecommand \@ifx [1]{%
 \ifx #1\expandafter \@firstoftwo
 \else \expandafter \@secondoftwo
 \fi
}%
\providecommand \natexlab [1]{#1}%
\providecommand \enquote  [1]{``#1''}%
\providecommand \bibnamefont  [1]{#1}%
\providecommand \bibfnamefont [1]{#1}%
\providecommand \citenamefont [1]{#1}%
\providecommand \href@noop [0]{\@secondoftwo}%
\providecommand \href [0]{\begingroup \@sanitize@url \@href}%
\providecommand \@href[1]{\@@startlink{#1}\@@href}%
\providecommand \@@href[1]{\endgroup#1\@@endlink}%
\providecommand \@sanitize@url [0]{\catcode `\\12\catcode `\$12\catcode
  `\&12\catcode `\#12\catcode `\^12\catcode `\_12\catcode `\%12\relax}%
\providecommand \@@startlink[1]{}%
\providecommand \@@endlink[0]{}%
\providecommand \url  [0]{\begingroup\@sanitize@url \@url }%
\providecommand \@url [1]{\endgroup\@href {#1}{\urlprefix }}%
\providecommand \urlprefix  [0]{URL }%
\providecommand \Eprint [0]{\href }%
\providecommand \doibase [0]{http://dx.doi.org/}%
\providecommand \selectlanguage [0]{\@gobble}%
\providecommand \bibinfo  [0]{\@secondoftwo}%
\providecommand \bibfield  [0]{\@secondoftwo}%
\providecommand \translation [1]{[#1]}%
\providecommand \BibitemOpen [0]{}%
\providecommand \bibitemStop [0]{}%
\providecommand \bibitemNoStop [0]{.\EOS\space}%
\providecommand \EOS [0]{\spacefactor3000\relax}%
\providecommand \BibitemShut  [1]{\csname bibitem#1\endcsname}%
\let\auto@bib@innerbib\@empty
\bibitem [{\citenamefont {Ruess}\ \emph {et~al.}(2004)\citenamefont {Ruess},
  \citenamefont {Oberbeck}, \citenamefont {Simmons}, \citenamefont {Goh},
  \citenamefont {Hamilton}, \citenamefont {Hallam}, \citenamefont {Schofield},
  \citenamefont {Curson},\ and\ \citenamefont {Clark}}]{r04}%
  \BibitemOpen
  \bibfield  {author} {\bibinfo {author} {\bibfnamefont {F.~J.}\ \bibnamefont
  {Ruess}}, \bibinfo {author} {\bibfnamefont {L.}~\bibnamefont {Oberbeck}},
  \bibinfo {author} {\bibfnamefont {M.}~\bibnamefont {Simmons}}, \bibinfo
  {author} {\bibfnamefont {K.}~\bibnamefont {Goh}}, \bibinfo {author}
  {\bibfnamefont {A.}~\bibnamefont {Hamilton}}, \bibinfo {author}
  {\bibfnamefont {T.}~\bibnamefont {Hallam}}, \bibinfo {author} {\bibfnamefont
  {S.~R.}\ \bibnamefont {Schofield}}, \bibinfo {author} {\bibfnamefont {N.~J.}\
  \bibnamefont {Curson}}, \ and\ \bibinfo {author} {\bibfnamefont {R.~G.}\
  \bibnamefont {Clark}},\ }\href@noop {} {\bibfield  {journal} {\bibinfo
  {journal} {Nano Lett.}\ }\textbf {\bibinfo {volume} {4}},\ \bibinfo {pages}
  {1969} (\bibinfo {year} {2004})}\BibitemShut {NoStop}%
\bibitem [{\citenamefont {Shen}\ \emph {et~al.}(2004)\citenamefont {Shen},
  \citenamefont {Kline}, \citenamefont {Schenkel}, \citenamefont {Robinson},
  \citenamefont {Ji}, \citenamefont {Yang}, \citenamefont {Du},\ and\
  \citenamefont {Tucker}}]{s04}%
  \BibitemOpen
  \bibfield  {author} {\bibinfo {author} {\bibfnamefont {T.-C.}\ \bibnamefont
  {Shen}}, \bibinfo {author} {\bibfnamefont {J.~S.}\ \bibnamefont {Kline}},
  \bibinfo {author} {\bibfnamefont {T.}~\bibnamefont {Schenkel}}, \bibinfo
  {author} {\bibfnamefont {S.~J.}\ \bibnamefont {Robinson}}, \bibinfo {author}
  {\bibfnamefont {J.-Y.}\ \bibnamefont {Ji}}, \bibinfo {author} {\bibfnamefont
  {C.}~\bibnamefont {Yang}}, \bibinfo {author} {\bibfnamefont {R.-R.}\
  \bibnamefont {Du}}, \ and\ \bibinfo {author} {\bibfnamefont {J.~R.}\
  \bibnamefont {Tucker}},\ }\href@noop {} {\bibfield  {journal} {\bibinfo
  {journal} {J. Vac. Sci. Technol. B}\ }\textbf {\bibinfo {volume} {22}},\
  \bibinfo {pages} {3182} (\bibinfo {year} {2004})}\BibitemShut {NoStop}%
\bibitem [{\citenamefont {RuessŸ}\ \emph {et~al.}(2007)\citenamefont {RuessŸ},
  \citenamefont {Pok}, \citenamefont {Reusch}, \citenamefont {Butcher},
  \citenamefont {Goh}, \citenamefont {Oberbeck}, \citenamefont {Scappucci},
  \citenamefont {Hamilton},\ and\ \citenamefont {Simmons}}]{f7}%
  \BibitemOpen
  \bibfield  {author} {\bibinfo {author} {\bibfnamefont {F.~J.}\ \bibnamefont
  {RuessŸ}}, \bibinfo {author} {\bibfnamefont {W.}~\bibnamefont {Pok}},
  \bibinfo {author} {\bibfnamefont {T.~C.}\ \bibnamefont {Reusch}}, \bibinfo
  {author} {\bibfnamefont {M.}~\bibnamefont {Butcher}}, \bibinfo {author}
  {\bibfnamefont {K.~E.}\ \bibnamefont {Goh}}, \bibinfo {author} {\bibfnamefont
  {L.}~\bibnamefont {Oberbeck}}, \bibinfo {author} {\bibfnamefont
  {G.}~\bibnamefont {Scappucci}}, \bibinfo {author} {\bibfnamefont
  {A.}~\bibnamefont {Hamilton}}, \ and\ \bibinfo {author} {\bibfnamefont
  {M.}~\bibnamefont {Simmons}},\ }\href@noop {} {\bibfield  {journal} {\bibinfo
   {journal} {Small}\ }\textbf {\bibinfo {volume} {3}},\ \bibinfo {pages} {563}
  (\bibinfo {year} {2007})}\BibitemShut {NoStop}%
\bibitem [{\citenamefont {Fuechsle}\ \emph {et~al.}(2012)\citenamefont
  {Fuechsle}, \citenamefont {Miwa}, \citenamefont {Mahapatra}, \citenamefont
  {Ryu}, \citenamefont {Lee}, \citenamefont {Warschkow}, \citenamefont
  {Hollenberg}, \citenamefont {Klimeck},\ and\ \citenamefont {Simmons}}]{f12}%
  \BibitemOpen
  \bibfield  {author} {\bibinfo {author} {\bibfnamefont {M.}~\bibnamefont
  {Fuechsle}}, \bibinfo {author} {\bibfnamefont {J.~A.}\ \bibnamefont {Miwa}},
  \bibinfo {author} {\bibfnamefont {S.}~\bibnamefont {Mahapatra}}, \bibinfo
  {author} {\bibfnamefont {H.}~\bibnamefont {Ryu}}, \bibinfo {author}
  {\bibfnamefont {S.}~\bibnamefont {Lee}}, \bibinfo {author} {\bibfnamefont
  {O.}~\bibnamefont {Warschkow}}, \bibinfo {author} {\bibfnamefont {L.~C.}\
  \bibnamefont {Hollenberg}}, \bibinfo {author} {\bibfnamefont
  {G.}~\bibnamefont {Klimeck}}, \ and\ \bibinfo {author} {\bibfnamefont
  {M.~Y.}\ \bibnamefont {Simmons}},\ }\href@noop {} {\bibfield  {journal}
  {\bibinfo  {journal} {Nat. Nanotech.}\ }\textbf {\bibinfo {volume} {7}},\
  \bibinfo {pages} {242} (\bibinfo {year} {2012})}\BibitemShut {NoStop}%
\bibitem [{\citenamefont {Scappucci}\ \emph {et~al.}(2009)\citenamefont
  {Scappucci}, \citenamefont {Capellini}, \citenamefont {Lee},\ and\
  \citenamefont {Simmons}}]{gs9}%
  \BibitemOpen
  \bibfield  {author} {\bibinfo {author} {\bibfnamefont {G.}~\bibnamefont
  {Scappucci}}, \bibinfo {author} {\bibfnamefont {G.}~\bibnamefont
  {Capellini}}, \bibinfo {author} {\bibfnamefont {W.~C.~T.}\ \bibnamefont
  {Lee}}, \ and\ \bibinfo {author} {\bibfnamefont {M.~Y.}\ \bibnamefont
  {Simmons}},\ }\href@noop {} {\bibfield  {journal} {\bibinfo  {journal} {Appl.
  Phys. Lett.}\ }\textbf {\bibinfo {volume} {94}},\ \bibinfo {pages} {162106}
  (\bibinfo {year} {2009})}\BibitemShut {NoStop}%
\bibitem [{\citenamefont {Scappucci}\ \emph {et~al.}(2011)\citenamefont
  {Scappucci}, \citenamefont {Capellini}, \citenamefont {Johnston},
  \citenamefont {Klesse}, \citenamefont {Miwa},\ and\ \citenamefont
  {Simmons}}]{gs11}%
  \BibitemOpen
  \bibfield  {author} {\bibinfo {author} {\bibfnamefont {G.}~\bibnamefont
  {Scappucci}}, \bibinfo {author} {\bibfnamefont {G.}~\bibnamefont
  {Capellini}}, \bibinfo {author} {\bibfnamefont {B.}~\bibnamefont {Johnston}},
  \bibinfo {author} {\bibfnamefont {W.~M.}\ \bibnamefont {Klesse}}, \bibinfo
  {author} {\bibfnamefont {J.~A.}\ \bibnamefont {Miwa}}, \ and\ \bibinfo
  {author} {\bibfnamefont {M.~Y.}\ \bibnamefont {Simmons}},\ }\href@noop {}
  {\bibfield  {journal} {\bibinfo  {journal} {Nano Lett.}\ }\textbf {\bibinfo
  {volume} {11}},\ \bibinfo {pages} {2272} (\bibinfo {year}
  {2011})}\BibitemShut {NoStop}%
\bibitem [{\citenamefont {Zwanenburg}\ \emph {et~al.}(2013)\citenamefont
  {Zwanenburg}, \citenamefont {Dzurak}, \citenamefont {Morello}, \citenamefont
  {Simmons}, \citenamefont {Hollenberg}, \citenamefont {Klimeck}, \citenamefont
  {Rogge}, \citenamefont {Coppersmith},\ and\ \citenamefont {Eriksson}}]{z12}%
  \BibitemOpen
  \bibfield  {author} {\bibinfo {author} {\bibfnamefont {F.~A.}\ \bibnamefont
  {Zwanenburg}}, \bibinfo {author} {\bibfnamefont {A.~S.}\ \bibnamefont
  {Dzurak}}, \bibinfo {author} {\bibfnamefont {A.}~\bibnamefont {Morello}},
  \bibinfo {author} {\bibfnamefont {M.~Y.}\ \bibnamefont {Simmons}}, \bibinfo
  {author} {\bibfnamefont {L.~C.~L.}\ \bibnamefont {Hollenberg}}, \bibinfo
  {author} {\bibfnamefont {G.}~\bibnamefont {Klimeck}}, \bibinfo {author}
  {\bibfnamefont {S.}~\bibnamefont {Rogge}}, \bibinfo {author} {\bibfnamefont
  {S.~N.}\ \bibnamefont {Coppersmith}}, \ and\ \bibinfo {author} {\bibfnamefont
  {M.~A.}\ \bibnamefont {Eriksson}},\ }\href {\doibase
  10.1103/RevModPhys.85.961} {\bibfield  {journal} {\bibinfo  {journal} {Rev.
  Mod. Phys.}\ }\textbf {\bibinfo {volume} {85}},\ \bibinfo {pages} {961}
  (\bibinfo {year} {2013})}\BibitemShut {NoStop}%
\bibitem [{\citenamefont {Kane}(1998)}]{K98}%
  \BibitemOpen
  \bibfield  {author} {\bibinfo {author} {\bibfnamefont {B.~E.}\ \bibnamefont
  {Kane}},\ }\href@noop {} {\bibfield  {journal} {\bibinfo  {journal} {Nature}\
  }\textbf {\bibinfo {volume} {393}},\ \bibinfo {pages} {133} (\bibinfo {year}
  {1998})}\BibitemShut {NoStop}%
\bibitem [{\citenamefont {Vrijen}\ \emph {et~al.}(2000)\citenamefont {Vrijen},
  \citenamefont {Yablonovitch}, \citenamefont {Wang}, \citenamefont {Jiang},
  \citenamefont {Balandin}, \citenamefont {Roychowdhury}, \citenamefont {Mor},\
  and\ \citenamefont {DiVincenzo}}]{Vr00}%
  \BibitemOpen
  \bibfield  {author} {\bibinfo {author} {\bibfnamefont {R.}~\bibnamefont
  {Vrijen}}, \bibinfo {author} {\bibfnamefont {E.}~\bibnamefont
  {Yablonovitch}}, \bibinfo {author} {\bibfnamefont {K.}~\bibnamefont {Wang}},
  \bibinfo {author} {\bibfnamefont {H.~W.}\ \bibnamefont {Jiang}}, \bibinfo
  {author} {\bibfnamefont {A.}~\bibnamefont {Balandin}}, \bibinfo {author}
  {\bibfnamefont {V.}~\bibnamefont {Roychowdhury}}, \bibinfo {author}
  {\bibfnamefont {T.}~\bibnamefont {Mor}}, \ and\ \bibinfo {author}
  {\bibfnamefont {D.}~\bibnamefont {DiVincenzo}},\ }\href {\doibase
  10.1103/PhysRevA.62.012306} {\bibfield  {journal} {\bibinfo  {journal} {Phys.
  Rev. A}\ }\textbf {\bibinfo {volume} {62}},\ \bibinfo {pages} {012306}
  (\bibinfo {year} {2000})}\BibitemShut {NoStop}%
\bibitem [{\citenamefont {Fang}\ \emph {et~al.}(2005)\citenamefont {Fang},
  \citenamefont {Chang},\ and\ \citenamefont {Tucker}}]{Fang05}%
  \BibitemOpen
  \bibfield  {author} {\bibinfo {author} {\bibfnamefont {A.}~\bibnamefont
  {Fang}}, \bibinfo {author} {\bibfnamefont {Y.-C.}\ \bibnamefont {Chang}}, \
  and\ \bibinfo {author} {\bibfnamefont {J.~R.}\ \bibnamefont {Tucker}},\
  }\href {\doibase 10.1103/PhysRevB.72.075355} {\bibfield  {journal} {\bibinfo
  {journal} {Phys. Rev. B}\ }\textbf {\bibinfo {volume} {72}},\ \bibinfo
  {pages} {075355} (\bibinfo {year} {2005})}\BibitemShut {NoStop}%
\bibitem [{\citenamefont {Hollenberg}\ \emph {et~al.}(2006)\citenamefont
  {Hollenberg}, \citenamefont {Greentree}, \citenamefont {Fowler},\ and\
  \citenamefont {Wellard}}]{H6}%
  \BibitemOpen
  \bibfield  {author} {\bibinfo {author} {\bibfnamefont {L.~C.~L.}\
  \bibnamefont {Hollenberg}}, \bibinfo {author} {\bibfnamefont {A.~D.}\
  \bibnamefont {Greentree}}, \bibinfo {author} {\bibfnamefont {A.~G.}\
  \bibnamefont {Fowler}}, \ and\ \bibinfo {author} {\bibfnamefont {C.~J.}\
  \bibnamefont {Wellard}},\ }\href {\doibase 10.1103/PhysRevB.74.045311}
  {\bibfield  {journal} {\bibinfo  {journal} {Phys. Rev. B}\ }\textbf {\bibinfo
  {volume} {74}},\ \bibinfo {pages} {045311} (\bibinfo {year}
  {2006})}\BibitemShut {NoStop}%
\bibitem [{\citenamefont {Hill}\ \emph {et~al.}(2015)\citenamefont {Hill},
  \citenamefont {Peretz}, \citenamefont {Hile}, \citenamefont {House},
  \citenamefont {Fuechsle}, \citenamefont {Rogge}, \citenamefont {Simmons},\
  and\ \citenamefont {Hollenberg}}]{H15}%
  \BibitemOpen
  \bibfield  {author} {\bibinfo {author} {\bibfnamefont {C.~D.}\ \bibnamefont
  {Hill}}, \bibinfo {author} {\bibfnamefont {E.}~\bibnamefont {Peretz}},
  \bibinfo {author} {\bibfnamefont {S.~J.}\ \bibnamefont {Hile}}, \bibinfo
  {author} {\bibfnamefont {M.~G.}\ \bibnamefont {House}}, \bibinfo {author}
  {\bibfnamefont {M.}~\bibnamefont {Fuechsle}}, \bibinfo {author}
  {\bibfnamefont {S.}~\bibnamefont {Rogge}}, \bibinfo {author} {\bibfnamefont
  {M.~Y.}\ \bibnamefont {Simmons}}, \ and\ \bibinfo {author} {\bibfnamefont
  {L.~C.~L.}\ \bibnamefont {Hollenberg}},\ }\href@noop {} {\bibfield  {journal}
  {\bibinfo  {journal} {Sci. Adv.}\ }\textbf {\bibinfo {volume} {1}},\ \bibinfo
  {eid} {e1500707} (\bibinfo {year} {2015})}\BibitemShut {NoStop}%
\bibitem [{\citenamefont {Koiller}\ \emph {et~al.}(2001)\citenamefont
  {Koiller}, \citenamefont {Hu},\ and\ \citenamefont {Das~Sarma}}]{k01}%
  \BibitemOpen
  \bibfield  {author} {\bibinfo {author} {\bibfnamefont {B.}~\bibnamefont
  {Koiller}}, \bibinfo {author} {\bibfnamefont {X.}~\bibnamefont {Hu}}, \ and\
  \bibinfo {author} {\bibfnamefont {S.}~\bibnamefont {Das~Sarma}},\ }\href
  {\doibase 10.1103/PhysRevLett.88.027903} {\bibfield  {journal} {\bibinfo
  {journal} {Phys. Rev. Lett.}\ }\textbf {\bibinfo {volume} {88}},\ \bibinfo
  {pages} {027903} (\bibinfo {year} {2001})}\BibitemShut {NoStop}%
\bibitem [{\citenamefont {Gamble}\ \emph {et~al.}(2015)\citenamefont {Gamble},
  \citenamefont {Jacobson}, \citenamefont {Nielsen}, \citenamefont {Baczewski},
  \citenamefont {Moussa}, \citenamefont {Monta\~no},\ and\ \citenamefont
  {Muller}}]{JK15}%
  \BibitemOpen
  \bibfield  {author} {\bibinfo {author} {\bibfnamefont {J.~K.}\ \bibnamefont
  {Gamble}}, \bibinfo {author} {\bibfnamefont {N.~T.}\ \bibnamefont
  {Jacobson}}, \bibinfo {author} {\bibfnamefont {E.}~\bibnamefont {Nielsen}},
  \bibinfo {author} {\bibfnamefont {A.~D.}\ \bibnamefont {Baczewski}}, \bibinfo
  {author} {\bibfnamefont {J.~E.}\ \bibnamefont {Moussa}}, \bibinfo {author}
  {\bibfnamefont {I.}~\bibnamefont {Monta\~no}}, \ and\ \bibinfo {author}
  {\bibfnamefont {R.~P.}\ \bibnamefont {Muller}},\ }\href {\doibase
  10.1103/PhysRevB.91.235318} {\bibfield  {journal} {\bibinfo  {journal} {Phys.
  Rev. B}\ }\textbf {\bibinfo {volume} {91}},\ \bibinfo {pages} {235318}
  (\bibinfo {year} {2015})}\BibitemShut {NoStop}%
\bibitem [{\citenamefont {Koenraad}\ and\ \citenamefont
  {Flatt\'e}(2011)}]{k11}%
  \BibitemOpen
  \bibfield  {author} {\bibinfo {author} {\bibfnamefont {P.}~\bibnamefont
  {Koenraad}}\ and\ \bibinfo {author} {\bibfnamefont {M.~E.}\ \bibnamefont
  {Flatt\'e}},\ }\href@noop {} {\bibfield  {journal} {\bibinfo  {journal} {Nat.
  Mater.}\ }\textbf {\bibinfo {volume} {10}},\ \bibinfo {pages} {91} (\bibinfo
  {year} {2011})}\BibitemShut {NoStop}%
\bibitem [{\citenamefont {Pok}\ \emph {et~al.}(2007)\citenamefont {Pok},
  \citenamefont {Reusch}, \citenamefont {Scappucci}, \citenamefont {Rueb},
  \citenamefont {Hamilton},\ and\ \citenamefont {Simmons}}]{P7}%
  \BibitemOpen
  \bibfield  {author} {\bibinfo {author} {\bibfnamefont {W.}~\bibnamefont
  {Pok}}, \bibinfo {author} {\bibfnamefont {T.~C.}\ \bibnamefont {Reusch}},
  \bibinfo {author} {\bibfnamefont {G.}~\bibnamefont {Scappucci}}, \bibinfo
  {author} {\bibfnamefont {F.~J.}\ \bibnamefont {Rueb}}, \bibinfo {author}
  {\bibfnamefont {A.~R.}\ \bibnamefont {Hamilton}}, \ and\ \bibinfo {author}
  {\bibfnamefont {M.~Y.}\ \bibnamefont {Simmons}},\ }\href@noop {} {\bibfield
  {journal} {\bibinfo  {journal} {IEEE Trans. Nanotechnol.}\ }\textbf {\bibinfo
  {volume} {6}},\ \bibinfo {pages} {213} (\bibinfo {year} {2007})}\BibitemShut
  {NoStop}%
\bibitem [{\citenamefont {Weber}\ \emph {et~al.}(2014)\citenamefont {Weber},
  \citenamefont {Ryu}, \citenamefont {Tan}, \citenamefont {Klimeck},\ and\
  \citenamefont {Simmons}}]{W14}%
  \BibitemOpen
  \bibfield  {author} {\bibinfo {author} {\bibfnamefont {B.}~\bibnamefont
  {Weber}}, \bibinfo {author} {\bibfnamefont {H.}~\bibnamefont {Ryu}}, \bibinfo
  {author} {\bibfnamefont {Y.-H.~M.}\ \bibnamefont {Tan}}, \bibinfo {author}
  {\bibfnamefont {G.}~\bibnamefont {Klimeck}}, \ and\ \bibinfo {author}
  {\bibfnamefont {M.~Y.}\ \bibnamefont {Simmons}},\ }\href@noop {} {\bibfield
  {journal} {\bibinfo  {journal} {Physical Review Letters}\ }\textbf {\bibinfo
  {volume} {113}},\ \bibinfo {pages} {246802} (\bibinfo {year}
  {2014})}\BibitemShut {NoStop}%
\bibitem [{\citenamefont {Scrymgeour}\ \emph {et~al.}(2017)\citenamefont
  {Scrymgeour}, \citenamefont {Baca}, \citenamefont {Fishgrab}, \citenamefont
  {Simonson}, \citenamefont {Marshall}, \citenamefont {Bussmann}, \citenamefont
  {Nakakura}, \citenamefont {Anderson},\ and\ \citenamefont {Misra}}]{Sc17}%
  \BibitemOpen
  \bibfield  {author} {\bibinfo {author} {\bibfnamefont {D.}~\bibnamefont
  {Scrymgeour}}, \bibinfo {author} {\bibfnamefont {A.}~\bibnamefont {Baca}},
  \bibinfo {author} {\bibfnamefont {K.}~\bibnamefont {Fishgrab}}, \bibinfo
  {author} {\bibfnamefont {R.}~\bibnamefont {Simonson}}, \bibinfo {author}
  {\bibfnamefont {M.}~\bibnamefont {Marshall}}, \bibinfo {author}
  {\bibfnamefont {E.}~\bibnamefont {Bussmann}}, \bibinfo {author}
  {\bibfnamefont {C.}~\bibnamefont {Nakakura}}, \bibinfo {author}
  {\bibfnamefont {M.}~\bibnamefont {Anderson}}, \ and\ \bibinfo {author}
  {\bibfnamefont {S.}~\bibnamefont {Misra}},\ }\href {\doibase
  http://dx.doi.org/10.1016/j.apsusc.2017.06.261} {\bibfield  {journal}
  {\bibinfo  {journal} {Applied Surface Science}\ }\textbf {\bibinfo {volume}
  {423}},\ \bibinfo {pages} {1097 } (\bibinfo {year} {2017})}\BibitemShut
  {NoStop}%
\bibitem [{\citenamefont {Gramse}\ \emph {et~al.}(2017)\citenamefont {Gramse},
  \citenamefont {K{\"o}lker}, \citenamefont {Lim}, \citenamefont {Stock},
  \citenamefont {Solanki}, \citenamefont {Schofield}, \citenamefont
  {Brinciotti}, \citenamefont {Aeppli}, \citenamefont {Kienberger},\ and\
  \citenamefont {Curson}}]{gr17}%
  \BibitemOpen
  \bibfield  {author} {\bibinfo {author} {\bibfnamefont {G.}~\bibnamefont
  {Gramse}}, \bibinfo {author} {\bibfnamefont {A.}~\bibnamefont {K{\"o}lker}},
  \bibinfo {author} {\bibfnamefont {T.}~\bibnamefont {Lim}}, \bibinfo {author}
  {\bibfnamefont {T.~J.}\ \bibnamefont {Stock}}, \bibinfo {author}
  {\bibfnamefont {H.}~\bibnamefont {Solanki}}, \bibinfo {author} {\bibfnamefont
  {S.~R.}\ \bibnamefont {Schofield}}, \bibinfo {author} {\bibfnamefont
  {E.}~\bibnamefont {Brinciotti}}, \bibinfo {author} {\bibfnamefont
  {G.}~\bibnamefont {Aeppli}}, \bibinfo {author} {\bibfnamefont
  {F.}~\bibnamefont {Kienberger}}, \ and\ \bibinfo {author} {\bibfnamefont
  {N.~J.}\ \bibnamefont {Curson}},\ }\href@noop {} {\bibfield  {journal}
  {\bibinfo  {journal} {Science Advances}\ }\textbf {\bibinfo {volume} {3}},\
  \bibinfo {pages} {e1602586} (\bibinfo {year} {2017})}\BibitemShut {NoStop}%
\bibitem [{\citenamefont {Scappucci}\ \emph {et~al.}(2007)\citenamefont
  {Scappucci}, \citenamefont {Ratto}, \citenamefont {Thompson}, \citenamefont
  {Reusch}, \citenamefont {Pok}, \citenamefont {Rue{\ss}}, \citenamefont
  {Rosei},\ and\ \citenamefont {Simmons}}]{S7}%
  \BibitemOpen
  \bibfield  {author} {\bibinfo {author} {\bibfnamefont {G.}~\bibnamefont
  {Scappucci}}, \bibinfo {author} {\bibfnamefont {F.}~\bibnamefont {Ratto}},
  \bibinfo {author} {\bibfnamefont {D.}~\bibnamefont {Thompson}}, \bibinfo
  {author} {\bibfnamefont {T.}~\bibnamefont {Reusch}}, \bibinfo {author}
  {\bibfnamefont {W.}~\bibnamefont {Pok}}, \bibinfo {author} {\bibfnamefont
  {F.}~\bibnamefont {Rue{\ss}}}, \bibinfo {author} {\bibfnamefont
  {F.}~\bibnamefont {Rosei}}, \ and\ \bibinfo {author} {\bibfnamefont
  {M.}~\bibnamefont {Simmons}},\ }\href@noop {} {\bibfield  {journal} {\bibinfo
   {journal} {Appl. Phys. Lett.}\ }\textbf {\bibinfo {volume} {91}},\ \bibinfo
  {pages} {222109} (\bibinfo {year} {2007})}\BibitemShut {NoStop}%
\bibitem [{\citenamefont {Fuechsle}\ \emph {et~al.}(2007)\citenamefont
  {Fuechsle}, \citenamefont {Rue{\ss}}, \citenamefont {Reusch}, \citenamefont
  {Mitic},\ and\ \citenamefont {Simmons}}]{MF7}%
  \BibitemOpen
  \bibfield  {author} {\bibinfo {author} {\bibfnamefont {M.}~\bibnamefont
  {Fuechsle}}, \bibinfo {author} {\bibfnamefont {F.~J.}\ \bibnamefont
  {Rue{\ss}}}, \bibinfo {author} {\bibfnamefont {T.~C.}\ \bibnamefont
  {Reusch}}, \bibinfo {author} {\bibfnamefont {M.}~\bibnamefont {Mitic}}, \
  and\ \bibinfo {author} {\bibfnamefont {M.~Y.}\ \bibnamefont {Simmons}},\
  }\href@noop {} {\bibfield  {journal} {\bibinfo  {journal} {J. Vac. Sci. and
  Technol. B}\ }\textbf {\bibinfo {volume} {25}},\ \bibinfo {pages} {2562}
  (\bibinfo {year} {2007})}\BibitemShut {NoStop}%
\bibitem [{\citenamefont {Lee}\ \emph {et~al.}(2010)\citenamefont {Lee},
  \citenamefont {Scappucci}, \citenamefont {Thompson},\ and\ \citenamefont
  {Simmons}}]{L10}%
  \BibitemOpen
  \bibfield  {author} {\bibinfo {author} {\bibfnamefont {W.}~\bibnamefont
  {Lee}}, \bibinfo {author} {\bibfnamefont {G.}~\bibnamefont {Scappucci}},
  \bibinfo {author} {\bibfnamefont {D.}~\bibnamefont {Thompson}}, \ and\
  \bibinfo {author} {\bibfnamefont {M.}~\bibnamefont {Simmons}},\ }\href@noop
  {} {\bibfield  {journal} {\bibinfo  {journal} {Appl. Phys. Lett.}\ }\textbf
  {\bibinfo {volume} {96}},\ \bibinfo {pages} {043116} (\bibinfo {year}
  {2010})}\BibitemShut {NoStop}%
\bibitem [{\citenamefont {Goh}\ \emph {et~al.}(2008)\citenamefont {Goh},
  \citenamefont {Augarten}, \citenamefont {Oberbeck},\ and\ \citenamefont
  {Simmons}}]{g9}%
  \BibitemOpen
  \bibfield  {author} {\bibinfo {author} {\bibfnamefont {K.~E.~J.}\
  \bibnamefont {Goh}}, \bibinfo {author} {\bibfnamefont {Y.}~\bibnamefont
  {Augarten}}, \bibinfo {author} {\bibfnamefont {L.}~\bibnamefont {Oberbeck}},
  \ and\ \bibinfo {author} {\bibfnamefont {M.~Y.}\ \bibnamefont {Simmons}},\
  }\href@noop {} {\bibfield  {journal} {\bibinfo  {journal} {Appl. Phys.
  Lett.}\ }\textbf {\bibinfo {volume} {93}},\ \bibinfo {eid} {142105} (\bibinfo
  {year} {2008})}\BibitemShut {NoStop}%
\bibitem [{\citenamefont {Sullivan}\ \emph {et~al.}(2004)\citenamefont
  {Sullivan}, \citenamefont {Kane},\ and\ \citenamefont {Thompson}}]{s4}%
  \BibitemOpen
  \bibfield  {author} {\bibinfo {author} {\bibfnamefont {D.~F.}\ \bibnamefont
  {Sullivan}}, \bibinfo {author} {\bibfnamefont {B.~E.}\ \bibnamefont {Kane}},
  \ and\ \bibinfo {author} {\bibfnamefont {P.~E.}\ \bibnamefont {Thompson}},\
  }\href@noop {} {\bibfield  {journal} {\bibinfo  {journal} {Appl. Phys.
  Lett.}\ }\textbf {\bibinfo {volume} {85}},\ \bibinfo {pages} {6362} (\bibinfo
  {year} {2004})}\BibitemShut {NoStop}%
\bibitem [{\citenamefont {Keizer}\ \emph {et~al.}(2015)\citenamefont {Keizer},
  \citenamefont {Koelling}, \citenamefont {Koenraad},\ and\ \citenamefont
  {Simmons}}]{k15}%
  \BibitemOpen
  \bibfield  {author} {\bibinfo {author} {\bibfnamefont {J.~G.}\ \bibnamefont
  {Keizer}}, \bibinfo {author} {\bibfnamefont {S.}~\bibnamefont {Koelling}},
  \bibinfo {author} {\bibfnamefont {P.~M.}\ \bibnamefont {Koenraad}}, \ and\
  \bibinfo {author} {\bibfnamefont {M.}~\bibnamefont {Simmons}},\ }\href@noop
  {} {\bibfield  {journal} {\bibinfo  {journal} {ACS Nano}\ }\textbf {\bibinfo
  {volume} {9}},\ \bibinfo {pages} {12537} (\bibinfo {year}
  {2015})}\BibitemShut {NoStop}%
\bibitem [{\citenamefont {Lee}\ \emph {et~al.}(2005)\citenamefont {Lee},
  \citenamefont {Fitzgerald}, \citenamefont {Bulsara}, \citenamefont {Currie},\
  and\ \citenamefont {Lochtefeld}}]{ML5}%
  \BibitemOpen
  \bibfield  {author} {\bibinfo {author} {\bibfnamefont {M.~L.}\ \bibnamefont
  {Lee}}, \bibinfo {author} {\bibfnamefont {E.~A.}\ \bibnamefont {Fitzgerald}},
  \bibinfo {author} {\bibfnamefont {M.~T.}\ \bibnamefont {Bulsara}}, \bibinfo
  {author} {\bibfnamefont {M.~T.}\ \bibnamefont {Currie}}, \ and\ \bibinfo
  {author} {\bibfnamefont {A.}~\bibnamefont {Lochtefeld}},\ }\href@noop {}
  {\bibfield  {journal} {\bibinfo  {journal} {J. Appl. Phys.}\ }\textbf
  {\bibinfo {volume} {97}},\ \bibinfo {eid} {011101} (\bibinfo {year}
  {2005})}\BibitemShut {NoStop}%
\bibitem [{\citenamefont {Friesen}\ \emph {et~al.}(2003)\citenamefont
  {Friesen}, \citenamefont {Rugheimer}, \citenamefont {Savage}, \citenamefont
  {Lagally}, \citenamefont {van~der Weide}, \citenamefont {Joynt},\ and\
  \citenamefont {Eriksson}}]{F3}%
  \BibitemOpen
  \bibfield  {author} {\bibinfo {author} {\bibfnamefont {M.}~\bibnamefont
  {Friesen}}, \bibinfo {author} {\bibfnamefont {P.}~\bibnamefont {Rugheimer}},
  \bibinfo {author} {\bibfnamefont {D.~E.}\ \bibnamefont {Savage}}, \bibinfo
  {author} {\bibfnamefont {M.~G.}\ \bibnamefont {Lagally}}, \bibinfo {author}
  {\bibfnamefont {D.~W.}\ \bibnamefont {van~der Weide}}, \bibinfo {author}
  {\bibfnamefont {R.}~\bibnamefont {Joynt}}, \ and\ \bibinfo {author}
  {\bibfnamefont {M.~A.}\ \bibnamefont {Eriksson}},\ }\href {\doibase
  10.1103/PhysRevB.67.121301} {\bibfield  {journal} {\bibinfo  {journal} {Phys.
  Rev. B}\ }\textbf {\bibinfo {volume} {67}},\ \bibinfo {pages} {121301}
  (\bibinfo {year} {2003})}\BibitemShut {NoStop}%
\bibitem [{\citenamefont {Pudalov}\ \emph {et~al.}(1993)\citenamefont
  {Pudalov}, \citenamefont {D'Iorio}, \citenamefont {Kravchenko},\ and\
  \citenamefont {Campbell}}]{P93}%
  \BibitemOpen
  \bibfield  {author} {\bibinfo {author} {\bibfnamefont {V.~M.}\ \bibnamefont
  {Pudalov}}, \bibinfo {author} {\bibfnamefont {M.}~\bibnamefont {D'Iorio}},
  \bibinfo {author} {\bibfnamefont {S.~V.}\ \bibnamefont {Kravchenko}}, \ and\
  \bibinfo {author} {\bibfnamefont {J.~W.}\ \bibnamefont {Campbell}},\ }\href
  {\doibase 10.1103/PhysRevLett.70.1866} {\bibfield  {journal} {\bibinfo
  {journal} {Phys. Rev. Lett.}\ }\textbf {\bibinfo {volume} {70}},\ \bibinfo
  {pages} {1866} (\bibinfo {year} {1993})}\BibitemShut {NoStop}%
\bibitem [{\citenamefont {Kim}\ \emph {et~al.}(2017)\citenamefont {Kim},
  \citenamefont {Tyryshkin},\ and\ \citenamefont {Lyon}}]{K17}%
  \BibitemOpen
  \bibfield  {author} {\bibinfo {author} {\bibfnamefont {J.-S.}\ \bibnamefont
  {Kim}}, \bibinfo {author} {\bibfnamefont {A.~M.}\ \bibnamefont {Tyryshkin}},
  \ and\ \bibinfo {author} {\bibfnamefont {S.~A.}\ \bibnamefont {Lyon}},\
  }\href@noop {} {\bibfield  {journal} {\bibinfo  {journal} {Applied Physics
  Letters}\ }\textbf {\bibinfo {volume} {110}},\ \bibinfo {pages} {123505}
  (\bibinfo {year} {2017})}\BibitemShut {NoStop}%
\bibitem [{\citenamefont {Lai}\ \emph {et~al.}(2005)\citenamefont {Lai},
  \citenamefont {Pan}, \citenamefont {Tsui}, \citenamefont {Lyon},
  \citenamefont {M\"uhlberger},\ and\ \citenamefont {Sch\"affler}}]{L05}%
  \BibitemOpen
  \bibfield  {author} {\bibinfo {author} {\bibfnamefont {K.}~\bibnamefont
  {Lai}}, \bibinfo {author} {\bibfnamefont {W.}~\bibnamefont {Pan}}, \bibinfo
  {author} {\bibfnamefont {D.~C.}\ \bibnamefont {Tsui}}, \bibinfo {author}
  {\bibfnamefont {S.~A.}\ \bibnamefont {Lyon}}, \bibinfo {author}
  {\bibfnamefont {M.}~\bibnamefont {M\"uhlberger}}, \ and\ \bibinfo {author}
  {\bibfnamefont {F.}~\bibnamefont {Sch\"affler}},\ }\href {\doibase
  10.1103/PhysRevB.72.081313} {\bibfield  {journal} {\bibinfo  {journal} {Phys.
  Rev. B}\ }\textbf {\bibinfo {volume} {72}},\ \bibinfo {pages} {081313}
  (\bibinfo {year} {2005})}\BibitemShut {NoStop}%
\bibitem [{\citenamefont {Lu}\ \emph {et~al.}(2009)\citenamefont {Lu},
  \citenamefont {Tsui}, \citenamefont {Lee},\ and\ \citenamefont {Liu}}]{l9}%
  \BibitemOpen
  \bibfield  {author} {\bibinfo {author} {\bibfnamefont {T.~M.}\ \bibnamefont
  {Lu}}, \bibinfo {author} {\bibfnamefont {D.~C.}\ \bibnamefont {Tsui}},
  \bibinfo {author} {\bibfnamefont {C.-H.}\ \bibnamefont {Lee}}, \ and\
  \bibinfo {author} {\bibfnamefont {C.~W.}\ \bibnamefont {Liu}},\ }\href@noop
  {} {\bibfield  {journal} {\bibinfo  {journal} {Appl. Phys. Lett.}\ }\textbf
  {\bibinfo {volume} {94}},\ \bibinfo {eid} {182102} (\bibinfo {year}
  {2009})}\BibitemShut {NoStop}%
\bibitem [{\citenamefont {Laroche}\ \emph {et~al.}(2015)\citenamefont
  {Laroche}, \citenamefont {Huang}, \citenamefont {Nielsen}, \citenamefont
  {Chuang}, \citenamefont {Li}, \citenamefont {Liu},\ and\ \citenamefont
  {Lu}}]{L15}%
  \BibitemOpen
  \bibfield  {author} {\bibinfo {author} {\bibfnamefont {D.}~\bibnamefont
  {Laroche}}, \bibinfo {author} {\bibfnamefont {S.-H.}\ \bibnamefont {Huang}},
  \bibinfo {author} {\bibfnamefont {E.}~\bibnamefont {Nielsen}}, \bibinfo
  {author} {\bibfnamefont {Y.}~\bibnamefont {Chuang}}, \bibinfo {author}
  {\bibfnamefont {J.-Y.}\ \bibnamefont {Li}}, \bibinfo {author} {\bibfnamefont
  {C.~W.}\ \bibnamefont {Liu}}, \ and\ \bibinfo {author} {\bibfnamefont
  {T.~M.}\ \bibnamefont {Lu}},\ }\href@noop {} {\bibfield  {journal} {\bibinfo
  {journal} {AIP Adv.}\ }\textbf {\bibinfo {volume} {5}},\ \bibinfo {eid}
  {107106} (\bibinfo {year} {2015})}\BibitemShut {NoStop}%
\bibitem [{\citenamefont {Mi}\ \emph {et~al.}(2015)\citenamefont {Mi},
  \citenamefont {Hazard}, \citenamefont {Payette}, \citenamefont {Wang},
  \citenamefont {Zajac}, \citenamefont {Cady},\ and\ \citenamefont
  {Petta}}]{XMi15}%
  \BibitemOpen
  \bibfield  {author} {\bibinfo {author} {\bibfnamefont {X.}~\bibnamefont
  {Mi}}, \bibinfo {author} {\bibfnamefont {T.~M.}\ \bibnamefont {Hazard}},
  \bibinfo {author} {\bibfnamefont {C.}~\bibnamefont {Payette}}, \bibinfo
  {author} {\bibfnamefont {K.}~\bibnamefont {Wang}}, \bibinfo {author}
  {\bibfnamefont {D.~M.}\ \bibnamefont {Zajac}}, \bibinfo {author}
  {\bibfnamefont {J.~V.}\ \bibnamefont {Cady}}, \ and\ \bibinfo {author}
  {\bibfnamefont {J.~R.}\ \bibnamefont {Petta}},\ }\href {\doibase
  10.1103/PhysRevB.92.035304} {\bibfield  {journal} {\bibinfo  {journal} {Phys.
  Rev. B}\ }\textbf {\bibinfo {volume} {92}},\ \bibinfo {pages} {035304}
  (\bibinfo {year} {2015})}\BibitemShut {NoStop}%
\bibitem [{\citenamefont {Zajac}\ \emph {et~al.}(2015)\citenamefont {Zajac},
  \citenamefont {Hazard}, \citenamefont {Mi}, \citenamefont {Wang},\ and\
  \citenamefont {Petta}}]{Z15}%
  \BibitemOpen
  \bibfield  {author} {\bibinfo {author} {\bibfnamefont {D.}~\bibnamefont
  {Zajac}}, \bibinfo {author} {\bibfnamefont {T.}~\bibnamefont {Hazard}},
  \bibinfo {author} {\bibfnamefont {X.}~\bibnamefont {Mi}}, \bibinfo {author}
  {\bibfnamefont {K.}~\bibnamefont {Wang}}, \ and\ \bibinfo {author}
  {\bibfnamefont {J.}~\bibnamefont {Petta}},\ }\href@noop {} {\bibfield
  {journal} {\bibinfo  {journal} {Applied Physics Letters}\ }\textbf {\bibinfo
  {volume} {106}},\ \bibinfo {pages} {223507} (\bibinfo {year}
  {2015})}\BibitemShut {NoStop}%
\bibitem [{\citenamefont {Klauk}\ \emph {et~al.}(1996)\citenamefont {Klauk},
  \citenamefont {Jackson}, \citenamefont {Nelson},\ and\ \citenamefont
  {Chu}}]{Kl96}%
  \BibitemOpen
  \bibfield  {author} {\bibinfo {author} {\bibfnamefont {H.}~\bibnamefont
  {Klauk}}, \bibinfo {author} {\bibfnamefont {T.~N.}\ \bibnamefont {Jackson}},
  \bibinfo {author} {\bibfnamefont {S.~F.}\ \bibnamefont {Nelson}}, \ and\
  \bibinfo {author} {\bibfnamefont {J.~O.}\ \bibnamefont {Chu}},\ }\href
  {\doibase 10.1063/1.115644} {\bibfield  {journal} {\bibinfo  {journal}
  {Applied Physics Letters}\ }\textbf {\bibinfo {volume} {68}},\ \bibinfo
  {pages} {1975} (\bibinfo {year} {1996})},\ \Eprint
  {http://arxiv.org/abs/http://dx.doi.org/10.1063/1.115644}
  {http://dx.doi.org/10.1063/1.115644} \BibitemShut {NoStop}%
\bibitem [{\citenamefont {Oberbeck}\ \emph {et~al.}(2002)\citenamefont
  {Oberbeck}, \citenamefont {Curson}, \citenamefont {Simmons}, \citenamefont
  {Brenner}, \citenamefont {Hamilton}, \citenamefont {Schofield},\ and\
  \citenamefont {Clark}}]{O2}%
  \BibitemOpen
  \bibfield  {author} {\bibinfo {author} {\bibfnamefont {L.}~\bibnamefont
  {Oberbeck}}, \bibinfo {author} {\bibfnamefont {N.}~\bibnamefont {Curson}},
  \bibinfo {author} {\bibfnamefont {M.}~\bibnamefont {Simmons}}, \bibinfo
  {author} {\bibfnamefont {R.}~\bibnamefont {Brenner}}, \bibinfo {author}
  {\bibfnamefont {A.}~\bibnamefont {Hamilton}}, \bibinfo {author}
  {\bibfnamefont {S.}~\bibnamefont {Schofield}}, \ and\ \bibinfo {author}
  {\bibfnamefont {R.}~\bibnamefont {Clark}},\ }\href@noop {} {\bibfield
  {journal} {\bibinfo  {journal} {Appl. Phys. Lett.}\ }\textbf {\bibinfo
  {volume} {81}},\ \bibinfo {pages} {3197} (\bibinfo {year}
  {2002})}\BibitemShut {NoStop}%
\bibitem [{\citenamefont {Lee}\ \emph {et~al.}(2013)\citenamefont {Lee},
  \citenamefont {Bishop}, \citenamefont {Thompson}, \citenamefont {Xue},
  \citenamefont {Scappucci}, \citenamefont {Cederberg}, \citenamefont {Gray},
  \citenamefont {Han}, \citenamefont {Celler}, \citenamefont {Carroll} \emph
  {et~al.}}]{l13}%
  \BibitemOpen
  \bibfield  {author} {\bibinfo {author} {\bibfnamefont {W.}~\bibnamefont
  {Lee}}, \bibinfo {author} {\bibfnamefont {N.}~\bibnamefont {Bishop}},
  \bibinfo {author} {\bibfnamefont {D.}~\bibnamefont {Thompson}}, \bibinfo
  {author} {\bibfnamefont {K.}~\bibnamefont {Xue}}, \bibinfo {author}
  {\bibfnamefont {G.}~\bibnamefont {Scappucci}}, \bibinfo {author}
  {\bibfnamefont {J.}~\bibnamefont {Cederberg}}, \bibinfo {author}
  {\bibfnamefont {J.}~\bibnamefont {Gray}}, \bibinfo {author} {\bibfnamefont
  {S.}~\bibnamefont {Han}}, \bibinfo {author} {\bibfnamefont {G.}~\bibnamefont
  {Celler}}, \bibinfo {author} {\bibfnamefont {M.}~\bibnamefont {Carroll}},
  \emph {et~al.},\ }\href@noop {} {\bibfield  {journal} {\bibinfo  {journal}
  {Appl. Surf. Sci.}\ }\textbf {\bibinfo {volume} {265}},\ \bibinfo {pages}
  {833} (\bibinfo {year} {2013})}\BibitemShut {NoStop}%
\bibitem [{\citenamefont {Lee}\ \emph {et~al.}(2014)\citenamefont {Lee},
  \citenamefont {McKibbin}, \citenamefont {Thompson}, \citenamefont {Xue},
  \citenamefont {Scappucci}, \citenamefont {Bishop}, \citenamefont {Celler},
  \citenamefont {Carroll},\ and\ \citenamefont {Simmons}}]{l14}%
  \BibitemOpen
  \bibfield  {author} {\bibinfo {author} {\bibfnamefont {W.}~\bibnamefont
  {Lee}}, \bibinfo {author} {\bibfnamefont {S.}~\bibnamefont {McKibbin}},
  \bibinfo {author} {\bibfnamefont {D.}~\bibnamefont {Thompson}}, \bibinfo
  {author} {\bibfnamefont {K.}~\bibnamefont {Xue}}, \bibinfo {author}
  {\bibfnamefont {G.}~\bibnamefont {Scappucci}}, \bibinfo {author}
  {\bibfnamefont {N.}~\bibnamefont {Bishop}}, \bibinfo {author} {\bibfnamefont
  {G.}~\bibnamefont {Celler}}, \bibinfo {author} {\bibfnamefont
  {M.}~\bibnamefont {Carroll}}, \ and\ \bibinfo {author} {\bibfnamefont
  {M.}~\bibnamefont {Simmons}},\ }\href@noop {} {\bibfield  {journal} {\bibinfo
   {journal} {Nanotechnology}\ }\textbf {\bibinfo {volume} {25}},\ \bibinfo
  {pages} {145302} (\bibinfo {year} {2014})}\BibitemShut {NoStop}%
\bibitem [{\citenamefont {Klesse}\ \emph {et~al.}(2013)\citenamefont {Klesse},
  \citenamefont {Scappucci}, \citenamefont {Capellini}, \citenamefont
  {Hartmann},\ and\ \citenamefont {Simmons}}]{k13}%
  \BibitemOpen
  \bibfield  {author} {\bibinfo {author} {\bibfnamefont {W.~M.}\ \bibnamefont
  {Klesse}}, \bibinfo {author} {\bibfnamefont {G.}~\bibnamefont {Scappucci}},
  \bibinfo {author} {\bibfnamefont {G.}~\bibnamefont {Capellini}}, \bibinfo
  {author} {\bibfnamefont {J.~M.}\ \bibnamefont {Hartmann}}, \ and\ \bibinfo
  {author} {\bibfnamefont {M.~Y.}\ \bibnamefont {Simmons}},\ }\href@noop {}
  {\bibfield  {journal} {\bibinfo  {journal} {Appl. Phys. Lett.}\ }\textbf
  {\bibinfo {volume} {102}},\ \bibinfo {eid} {151103} (\bibinfo {year}
  {2013})}\BibitemShut {NoStop}%
\bibitem [{\citenamefont {Ishizaka}\ and\ \citenamefont {Shiraki}(1986)}]{i86}%
  \BibitemOpen
  \bibfield  {author} {\bibinfo {author} {\bibfnamefont {A.}~\bibnamefont
  {Ishizaka}}\ and\ \bibinfo {author} {\bibfnamefont {Y.}~\bibnamefont
  {Shiraki}},\ }\href@noop {} {\bibfield  {journal} {\bibinfo  {journal} {J.
  Electrochem. Soc.Journal of the Electrochemical Society}\ }\textbf {\bibinfo
  {volume} {133}},\ \bibinfo {pages} {666} (\bibinfo {year}
  {1986})}\BibitemShut {NoStop}%
\bibitem [{\citenamefont {Okumura}\ \emph {et~al.}(1998)\citenamefont
  {Okumura}, \citenamefont {Akane},\ and\ \citenamefont {Matsumoto}}]{O98}%
  \BibitemOpen
  \bibfield  {author} {\bibinfo {author} {\bibfnamefont {H.}~\bibnamefont
  {Okumura}}, \bibinfo {author} {\bibfnamefont {T.}~\bibnamefont {Akane}}, \
  and\ \bibinfo {author} {\bibfnamefont {S.}~\bibnamefont {Matsumoto}},\
  }\href@noop {} {\bibfield  {journal} {\bibinfo  {journal} {Appl. Surf. Sci.}\
  }\textbf {\bibinfo {volume} {125}},\ \bibinfo {pages} {125} (\bibinfo {year}
  {1998})}\BibitemShut {NoStop}%
\bibitem [{\citenamefont {Tersoff}\ \emph {et~al.}(1995)\citenamefont
  {Tersoff}, \citenamefont {Phang}, \citenamefont {Zhang},\ and\ \citenamefont
  {Lagally}}]{T95}%
  \BibitemOpen
  \bibfield  {author} {\bibinfo {author} {\bibfnamefont {J.}~\bibnamefont
  {Tersoff}}, \bibinfo {author} {\bibfnamefont {Y.}~\bibnamefont {Phang}},
  \bibinfo {author} {\bibfnamefont {Z.}~\bibnamefont {Zhang}}, \ and\ \bibinfo
  {author} {\bibfnamefont {M.}~\bibnamefont {Lagally}},\ }\href@noop {}
  {\bibfield  {journal} {\bibinfo  {journal} {Phys. Rev. Lett.}\ }\textbf
  {\bibinfo {volume} {75}},\ \bibinfo {pages} {2730} (\bibinfo {year}
  {1995})}\BibitemShut {NoStop}%
\bibitem [{\citenamefont {McKibbin}\ \emph {et~al.}(2009)\citenamefont
  {McKibbin}, \citenamefont {Clarke}, \citenamefont {Fuhrer}, \citenamefont
  {Reusch},\ and\ \citenamefont {Simmons}}]{m9}%
  \BibitemOpen
  \bibfield  {author} {\bibinfo {author} {\bibfnamefont {S.}~\bibnamefont
  {McKibbin}}, \bibinfo {author} {\bibfnamefont {W.}~\bibnamefont {Clarke}},
  \bibinfo {author} {\bibfnamefont {A.}~\bibnamefont {Fuhrer}}, \bibinfo
  {author} {\bibfnamefont {T.}~\bibnamefont {Reusch}}, \ and\ \bibinfo {author}
  {\bibfnamefont {M.}~\bibnamefont {Simmons}},\ }\href@noop {} {\bibfield
  {journal} {\bibinfo  {journal} {Appl. Phys. Lett.}\ }\textbf {\bibinfo
  {volume} {95}},\ \bibinfo {pages} {233111} (\bibinfo {year}
  {2009})}\BibitemShut {NoStop}%
\bibitem [{\citenamefont {McKibbin}\ \emph {et~al.}(2013)\citenamefont
  {McKibbin}, \citenamefont {Scappucci}, \citenamefont {Pok},\ and\
  \citenamefont {Simmons}}]{m13}%
  \BibitemOpen
  \bibfield  {author} {\bibinfo {author} {\bibfnamefont {S.}~\bibnamefont
  {McKibbin}}, \bibinfo {author} {\bibfnamefont {G.}~\bibnamefont {Scappucci}},
  \bibinfo {author} {\bibfnamefont {W.}~\bibnamefont {Pok}}, \ and\ \bibinfo
  {author} {\bibfnamefont {M.}~\bibnamefont {Simmons}},\ }\href@noop {}
  {\bibfield  {journal} {\bibinfo  {journal} {Nanotechnology}\ }\textbf
  {\bibinfo {volume} {24}},\ \bibinfo {pages} {045303} (\bibinfo {year}
  {2013})}\BibitemShut {NoStop}%
\bibitem [{\citenamefont {Finch}\ \emph {et~al.}(1963)\citenamefont {Finch},
  \citenamefont {Queisser}, \citenamefont {Thomas},\ and\ \citenamefont
  {Washburn}}]{Fi63}%
  \BibitemOpen
  \bibfield  {author} {\bibinfo {author} {\bibfnamefont {R.~H.}\ \bibnamefont
  {Finch}}, \bibinfo {author} {\bibfnamefont {H.~J.}\ \bibnamefont {Queisser}},
  \bibinfo {author} {\bibfnamefont {G.}~\bibnamefont {Thomas}}, \ and\ \bibinfo
  {author} {\bibfnamefont {J.}~\bibnamefont {Washburn}},\ }\href {\doibase
  10.1063/1.1702622} {\bibfield  {journal} {\bibinfo  {journal} {Journal of
  Applied Physics}\ }\textbf {\bibinfo {volume} {34}},\ \bibinfo {pages} {406}
  (\bibinfo {year} {1963})},\ \Eprint
  {http://arxiv.org/abs/http://dx.doi.org/10.1063/1.1702622}
  {http://dx.doi.org/10.1063/1.1702622} \BibitemShut {NoStop}%
\bibitem [{\citenamefont {Hwang}\ and\ \citenamefont {Sarma}(2013)}]{H13}%
  \BibitemOpen
  \bibfield  {author} {\bibinfo {author} {\bibfnamefont {E.}~\bibnamefont
  {Hwang}}\ and\ \bibinfo {author} {\bibfnamefont {S.~D.}\ \bibnamefont
  {Sarma}},\ }\href@noop {} {\bibfield  {journal} {\bibinfo  {journal} {Phys.
  Rev. B}\ }\textbf {\bibinfo {volume} {87}},\ \bibinfo {pages} {125411}
  (\bibinfo {year} {2013})}\BibitemShut {NoStop}%
\bibitem [{\citenamefont {Carter}\ \emph {et~al.}(2009)\citenamefont {Carter},
  \citenamefont {Warschkow}, \citenamefont {Marks},\ and\ \citenamefont
  {McKenzie}}]{carter:2009}%
  \BibitemOpen
  \bibfield  {author} {\bibinfo {author} {\bibfnamefont {D.~J.}\ \bibnamefont
  {Carter}}, \bibinfo {author} {\bibfnamefont {O.}~\bibnamefont {Warschkow}},
  \bibinfo {author} {\bibfnamefont {N.~A.}\ \bibnamefont {Marks}}, \ and\
  \bibinfo {author} {\bibfnamefont {D.~R.}\ \bibnamefont {McKenzie}},\ }\href
  {\doibase 10.1103/PhysRevB.79.033204} {\bibfield  {journal} {\bibinfo
  {journal} {Phys. Rev. B}\ }\textbf {\bibinfo {volume} {79}},\ \bibinfo
  {pages} {033204} (\bibinfo {year} {2009})}\BibitemShut {NoStop}%
\bibitem [{\citenamefont {Trellakis}\ \emph {et~al.}(1997)\citenamefont
  {Trellakis}, \citenamefont {Galick}, \citenamefont {Pacelli},\ and\
  \citenamefont {Ravaioli}}]{trellakis:1997}%
  \BibitemOpen
  \bibfield  {author} {\bibinfo {author} {\bibfnamefont {A.}~\bibnamefont
  {Trellakis}}, \bibinfo {author} {\bibfnamefont {A.}~\bibnamefont {Galick}},
  \bibinfo {author} {\bibfnamefont {A.}~\bibnamefont {Pacelli}}, \ and\
  \bibinfo {author} {\bibfnamefont {U.}~\bibnamefont {Ravaioli}},\ }\href@noop
  {} {\bibfield  {journal} {\bibinfo  {journal} {J. Appl. Phys.}\ }\textbf
  {\bibinfo {volume} {81}},\ \bibinfo {pages} {7880} (\bibinfo {year}
  {1997})}\BibitemShut {NoStop}%
\bibitem [{\citenamefont {Gao}\ \emph {et~al.}(2013)\citenamefont {Gao},
  \citenamefont {Nielsen}, \citenamefont {Muller}, \citenamefont {Young},
  \citenamefont {Salinger}, \citenamefont {Bishop}, \citenamefont {Lilly},\
  and\ \citenamefont {Carroll}}]{gao:2013}%
  \BibitemOpen
  \bibfield  {author} {\bibinfo {author} {\bibfnamefont {X.}~\bibnamefont
  {Gao}}, \bibinfo {author} {\bibfnamefont {E.}~\bibnamefont {Nielsen}},
  \bibinfo {author} {\bibfnamefont {R.~P.}\ \bibnamefont {Muller}}, \bibinfo
  {author} {\bibfnamefont {R.~W.}\ \bibnamefont {Young}}, \bibinfo {author}
  {\bibfnamefont {A.~G.}\ \bibnamefont {Salinger}}, \bibinfo {author}
  {\bibfnamefont {N.~C.}\ \bibnamefont {Bishop}}, \bibinfo {author}
  {\bibfnamefont {M.~P.}\ \bibnamefont {Lilly}}, \ and\ \bibinfo {author}
  {\bibfnamefont {M.~S.}\ \bibnamefont {Carroll}},\ }\href@noop {} {\bibfield
  {journal} {\bibinfo  {journal} {J. Appl. Phys.}\ }\textbf {\bibinfo {volume}
  {114}},\ \bibinfo {pages} {164302} (\bibinfo {year} {2013})}\BibitemShut
  {NoStop}%
\bibitem [{\citenamefont {Drumm}\ \emph {et~al.}(2012)\citenamefont {Drumm},
  \citenamefont {Hollenberg}, \citenamefont {Simmons},\ and\ \citenamefont
  {Friesen}}]{drumm:2012}%
  \BibitemOpen
  \bibfield  {author} {\bibinfo {author} {\bibfnamefont {D.~W.}\ \bibnamefont
  {Drumm}}, \bibinfo {author} {\bibfnamefont {L.~C.}\ \bibnamefont
  {Hollenberg}}, \bibinfo {author} {\bibfnamefont {M.~Y.}\ \bibnamefont
  {Simmons}}, \ and\ \bibinfo {author} {\bibfnamefont {M.}~\bibnamefont
  {Friesen}},\ }\href@noop {} {\bibfield  {journal} {\bibinfo  {journal} {Phys.
  Rev. B}\ }\textbf {\bibinfo {volume} {85}},\ \bibinfo {pages} {155419}
  (\bibinfo {year} {2012})}\BibitemShut {NoStop}%
\end{thebibliography}%

\end{document}